\documentclass[manuscript,screen,acmsmall]{acmart}

\AtBeginDocument{%
  }
\setcopyright{acmlicensed}
\copyrightyear{2018}
\acmYear{2018}
\acmDOI{XXXXXXX.XXXXXXX}
\acmConference[Conference acronym 'XX]{Make sure to enter the correct
  conference title from your rights confirmation email}{June 03--05,
  2018}{Woodstock, NY}
\acmISBN{978-1-4503-XXXX-X/2018/06}

\usepackage{lipsum}
\usepackage[textsize=tiny]{todonotes}
\usepackage{enumitem}
\usepackage{float}
\usepackage{longtable} 
\usepackage{adjustbox}
\usepackage{geometry}
\usepackage{threeparttable}
\usepackage{xcolor}
\usepackage{booktabs}
\usepackage{longtable}
\usepackage{pdflscape}
\usepackage{array}
\usepackage{ragged2e}

\newcolumntype{L}[1]{>{\RaggedRight\arraybackslash}p{#1}}

\begin{document}


\title{What Are Adversaries Doing? Automating Tactics, Techniques, and Procedures Extraction: A Systematic Review}

\author{Mahzabin Tamanna}
\affiliation{%
  \institution{North Carolina State University}
  \city{Raleigh}
  \state{NC}
  \country{USA}}
\email{mtamann@ncsu.edu}

\author{Shaswata Mitra}
\affiliation{%
  \institution{The University of Alabama}
  \city{Tuscaloosa}
  \state{AL}
  \country{USA}}
\email{smitra3@crimson.ua.edu}

\author{Md Erfan}
\affiliation{%
  \institution{The University of Alabama}
  \city{Tuscaloosa}
  \state{AL}
  \country{USA}}
\email{merfan@crimson.ua.edu}

\author{Ahmed Ryan}
\affiliation{%
  \institution{The University of Alabama}
  \city{Tuscaloosa}
  \state{AL}
  \country{USA}}
\email{aryan9@crimson.ua.edu}

\author{Sudip Mittal}
\affiliation{%
  \institution{The University of Alabama}
  \city{Tuscaloosa}
  \state{AL}
  \country{USA}}
  \email{sudip.mittal@ua.edu}

\author{Laurie Williams}
\affiliation{%
  \institution{North Carolina State University}
  \city{Raleigh}
  \state{NC}
  \country{USA}}
\email{lawilli3@ncsu.edu}

\author{Md Rayhanur Rahman}
\affiliation{%
  \institution{The University of Alabama}
  \city{Tuscaloosa}
  \state{AL}
  \country{USA}}
\email{mrahman87@ua.edu}

\renewcommand{\shortauthors}{Tamanna et al.}
\def\attack{ATT\&CK }

\begin{abstract}
 Adversaries continuously evolve their tactics, techniques, and procedures (TTPs) to achieve their objectives while evading detection, requiring defenders to continually update their understanding of adversary behavior. Prior research has proposed automated extraction of TTP-related intelligence from unstructured text and mapping it to structured knowledge bases, such as MITRE ATT\&CK. However, existing work varies widely in its extraction objectives, datasets, modeling approaches, and evaluation practices, making it difficult to understand the research landscape.
\textit{The goal of this study is to aid security researchers in understanding the state of the art in extracting attack tactics, techniques, and procedures (TTPs) from unstructured text by analyzing relevant literature.} We systematically analyze 80 peer-reviewed studies and categorize them across several key research dimensions, including extraction purposes, textual data sources, dataset construction and annotation strategies, modeling approaches, evaluation metrics, and artifact availability. In addition, our analysis reveals several dominant trends. Technique-level classification remains the dominant task formulation, while tactic classification and technique searching are comparatively underexplored. The field has progressed from rule-based and traditional machine learning approaches to transformer-based architectures (e.g., BERT, SecureBERT, RoBERTa), which improve contextual modeling of CTI text. More recently, studies have begun exploring Large Language Model (LLM) based approaches, including prompting, retrieval-augmented generation, and lightweight fine-tuning, though adoption remains emergent. Despite these advances, important limitations remain. Many studies rely on single-label or simplified classification settings, limited evaluation settings, and single or narrowly scoped datasets, which limit cross-domain generalization and realistic evaluation. Reproducibility is also constrained because many works rely on proprietary datasets, limited-release code, or annotated corpora, hindering independent validation and systematic benchmarking. Based on these findings, we outline future research directions to develop more robust, reproducible, and operationally relevant TTP extraction systems that better align with real-world CTI workflows.

\end{abstract}
\ccsdesc[500]{Computing methodologies~Information Retrieval}
\ccsdesc[300]{Security and Privacy}
\keywords{ Cyberthreat Intelligence, Tactics Techniques and Procedures (TTP) Extraction, MITRE ATT\&CK, Large Language Models (LLMs), Knowledge Graphs, Information Extraction}
\maketitle


\section{Introduction}
The growing scale and complexity of cyberattacks are posing serious challenges to the security and stability of organizations worldwide. Security breaches surged by nearly 75\% in 2024 compared to the previous year, with organizations experiencing an average of 1,876 cyberattacks per quarter~\cite{accn}. In 2025, the global average cost of a data breach reached \$4.44 million and \$10.22 million in the United States~\cite{IBM,veronis}. According to estimates by the International Monetary Fund (IMF), global cybercrime is projected to cost the world \$23 trillion by 2027, representing a 175\% increase compared to 2022~\cite{sentinelone}. The increase in volume and sophistication of threats has made it imperative for defenders to leverage timely Cyber Threat Intelligence (CTI). With the rapid increase in cyber incidents, the number of CTI reports published by security vendors and research communities has surged~\cite {liu2022threat}. The primary purpose of CTI is to understand the adversaries’ Tactics, Techniques, and Procedures (TTPs), where tactics represent the objectives of an attack (why), techniques describe the methods used to achieve those objectives (how), and procedures detail how adversaries implement those techniques in practice~\cite{mitre}. TTPs characterize how threat actors plan and execute cyberattacks, providing a structured foundation for analyzing adversarial behavior and for informing the design of defensive strategies. TTPs enable organizations to map real-world observations of adversarial activity to well-established attack patterns, such as those captured in the MITRE ATT\&CK framework. In parallel, TTPs support the systematic alignment of adversarial actions with defensive countermeasures via MITRE D3FEND~\cite{mitre_d3fend}, which complements attack-centric threat modeling. Mapping unstructured CTI reports to TTPs allows analysts to better interpret, correlate, and mitigate observed attacks. However, manual analysis and extraction of actionable insights from the rapidly growing volume of CTI reports is labor-intensive, error-prone, and does not scale with the accelerating pace of cyber threats~\cite{deliu2018collecting,gao2021enabling,mijwil2023significance}, which motivates research into automated TTP extraction from textual sources.

Several studies have proposed a variety of methods for automatic TTPs extraction, including Rule-Based and Pattern-Matching systems~\cite{hassan2020tactical, milajerdi2019holmes,zhu2023aptshield,rahman2024attackers}, Deep Learning models~\cite{lima2025toward,kumar2021dltif,behzadan2018corpus}, Natural Language Processing (NLP) pipelines~\cite{huang2024mitretrieval,orbinato2022automatic, gao2021enabling, fujii2023extracting}, and more recently, Large Language Model (LLM)-based approaches~\cite{chen2025aecr, fayyazi2024advancing, kim2025multi,zhang2024unittp,rahman2024towards}. These methods differ in their scope and granularity; some focus on high-level tactics, while others aim to identify fine-grained techniques or procedures. Existing works often rely on proprietary or non-public CTI datasets, use inconsistent definitions or ontologies of TTPs, and employ diverse evaluation metrics, which makes it difficult to compare results or assess generalizability. Consequently, remains unclear which types of TTP-related information are being extracted, which methodologies are most effective, and how these methods perform relative to each other. While a meta-study by Büchel et al.~\cite{buchel2025sok} has examined NLP-based solutions for knowledge extraction from CTI, there exists a gap in the synthesis of \textit{approaches, datasets, and evaluation} strategies for TTP extraction. To the best of our knowledge, no systematic literature review or knowledge synthesis has yet addressed the mentioned concerns. 


\textit{The goal of this study is to aid security researchers in understanding the state of the art of attack tactics, techniques, and procedures (TTPs) extraction from unstructured text through an analysis of relevant studies in the literature}. By consolidating and analyzing the diverse body of research on this topic, we provide a structured synthesis that will help the community understand the strengths and limitations of existing approaches, the sources and resources available, and identify where future efforts are needed. To achieve our goal, we answer the following research questions (RQs):

\begin{itemize}
    \item RQ1:What are the primary objectives of TTP-related information extraction?
    \item RQ2: What sources have been used for data collection?
    
    \item RQ3: How is the dataset collected and pre-processed?

    \item RQ4: How is the dataset annotated and constructed for the experiment?

     \item RQ5: What methodological approaches have been employed for TTPs extraction?

    \item RQ6: What metrics are used to evaluate performance?

     \item {RQ7: To what extent do existing studies provide reproducibility and replicability of TTP extraction systems?}



\end{itemize}

To answer these RQs, we systematically gathered 80 relevant studies focusing on the extraction of TTPs from textual descriptions through targeted keyword searches. We then conducted a qualitative analysis of these studies using open coding~\cite{saldana2021coding} and categorized the findings. Our key contributions are summarized as follows: (a) A systematic categorization of TTP extraction purposes, (b) A systematic categorization of methods associated with TTP extraction, (c) A set of data sources and formats utilized in the studies, (d) A compilation of approaches applied for evaluating the performance of the extract methods, (e) { A systematic overview of artifact accessibility, transparency, and reproducibility support within the existing body of work}, and  (f) A set of recommendations for security researchers for future research in TTP extraction.

\section{Key Concepts}\label{2}
\textbf{Tactics, Techniques, and Procedures (TTPs)}: TTPs (Tactics, Techniques, and Procedures) describe how threat actors plan and execute cyberattacks, helping defenders understand the why, what, and how of adversary behavior. A tactic captures the high‑level objective of an attack (e.g., persistence or exfiltration), explaining why an action is taken. A technique defines what method is used to achieve that objective, such as credential dumping or brute force. A procedure specifies how a technique is implemented in practice, including concrete steps, tools, and commands observed in real attacks. Together, TTPs enable more effective threat detection and mitigation through frameworks like MITRE ATT\&CK.

 \textbf{MITRE ATT\&CK}: The MITRE ATT\&CK taxonomy \cite{mitre} is a widely adopted knowledge base for cataloging adversary TTPs across the cyberattack lifecycle, providing a standardized, matrix‑based model of attacker behavior. It serves as a common language for modeling and communicating threat activity within the cybersecurity community. Complementing ATT\&CK, MITRE D3FEND\cite{mitre_d3fend} formalizes defensive countermeasures and their relationships to adversary techniques. Together, ATT\&CK and D3FEND form an integrated offense–defense ecosystem that links attack behaviors to corresponding defensive strategies, enabling more proactive, intelligence‑driven security planning and improved threat detection.


 \textbf{Cyber Threat Intelligence (CTI)}: Cyber Threat Intelligence (CTI) refers to analyzed and contextualized knowledge about cyber threats that supports informed decision‑making across cybersecurity operations. Unlike raw security data, CTI captures adversary context, including attacker identities, tactics, techniques, and procedures (TTPs), motivations, and observable indicators used for detection and response. By combining technical artifacts such as Indicators of Compromise (IoCs) with behavioral and strategic insights, CTI enables a shift from reactive incident response to proactive and anticipatory defense. CTI is typically derived from heterogeneous and largely unstructured sources, including threat reports, vulnerability disclosures, system logs, social media, and underground forums, and is organized using standardized frameworks such as MITRE ATT\&CK and STIX. As CTI supports functions ranging from detection and threat hunting to attribution and strategic risk assessment, automating its extraction and structuring from textual and operational data has become a central research challenge in cybersecurity.


 \textbf{Structured Threat Information Expression (STIX)}: Structured Threat Information Expression (STIX)~\cite{stix} is a standardized, machine‑readable language for representing and sharing cyber threat intelligence. It provides a unified vocabulary for describing threat actors, their activities, technical indicators, and defensive measures, as well as the relationships among them. STIX adopts a graph‑based data model in which threat entities are represented as nodes and their semantic relationships as edges, enabling flexible correlation and reasoning across heterogeneous CTI sources. This structure naturally aligns with knowledge bases such as MITRE ATT\&CK and MITRE D3FEND, which similarly organize adversary behaviors and defensive techniques as interconnected concepts.

 \textbf{Indicator of Compromise (IoC)}: An Indicator of Compromise (IoC) is an observable artifact that suggests a system or network has been affected by malicious activity. Common IoCs include malicious IP addresses, file hashes, abnormal network traffic patterns, and system or configuration changes. IoCs capture low‑level, procedural evidence of attacker behavior and support detection, correlation, and incident response. However, IoCs are often short‑lived due to attacker adaptation, automated artifact generation, and frequent changes in infrastructure~\cite{medium, NETRESEC2,cybereason}.


\section{Related Work}\label{3}
Several studies have explored the extraction of Cyber Threat Intelligence (CTI) from unstructured sources in the field of cybersecurity~\cite{rahman2020literature,rahman2024chronocti,ma2025ctiminer,wang2024knowcti}.
 Rahman et al.~\cite{rahman2023attackers} performed a systematic literature review of automated CTI extraction studies while focusing on the utilization of natural language processing (NLP) and machine learning (ML) techniques, where TTP extraction appears as one task among many.
Santos et al.~\cite{santos2025systematic} further synthesized CTI applications across sectors and highlighted challenges such as standardization, privacy, and trust, yet did not analyze the methodological nuances of ATT\&CK. Saeed~\cite{saeed2023systematic} examined CTI from an organizational resilience standpoint, proposing a layered framework of knowledge bases, detection models, and dashboards; nevertheless, their emphasis lies on CTI adoption and operational capability rather than the computational characteristics of TTP extraction. However, none of the studies provide a focused, in-depth synthesis of TTP extraction as a distinct research problem or provided detail insight.

Prior studies have also examined theoretical and representational aspects of CTI and examined how CTI is formally represented, structured, and modeled using frameworks. Roy et al.~\cite{roy2023sok} systematized knowledge around the MITRE ATT\&CK framework, demonstrating academia’s focus on NLP-based intelligence retrieval and industry’s reliance on ATT\&CK matrices for threat scoring. Bratsas et al.~\cite{bratsas2024knowledge} extended this line by reviewing knowledge graphs and semantic web ontologies for representing CTI, highlighting the growing importance of structured threat representation.
Bridges et al.~\cite{bridges2017cybersecurity} conducted a comparative study of prior cybersecurity entity extraction methods~\cite{bridges2013automatic,joshi2013extracting,jones2015towards}, utilizing data sources such as cybersecurity blogs, the National Vulnerability Database (NVD)~\cite{NVD}, and the Common Vulnerabilities and Exposures (CVE)~\cite{CVE} repositories. Their findings highlighted two critical limitations: low recall performance and the absence of publicly available datasets. Wang~\cite{wang2022systematic} reviewed the use of CTI in cyber threat hunting, exploring its applications in industrial control systems and next-generation security operations centers. While this work briefly mentions the use of TTPs in hunting activities, it does not examine automated TTP extraction methodologies or their technical evolution. Other studies investigated the technical and organizational aspects of CTI sharing. Tounsi et al. ~\cite{tounsi2018survey} classified CTI into four major categories and examined the landscape of technical CTI, identifying existing challenges, ongoing research trends, and the comparative features of current CTI collection and sharing tools. Similarly, Wagner et al.~\cite{wagner2019cyber} investigated both technical and nontechnical barriers to modern CTI sharing systems. Sauerwein et al. ~\cite{sauerwein2017threat} analyzed 22 CTI-sharing platforms, focusing on their ability to automate the generation, refinement, and analysis of threat intelligence data. Complementarily, Tuma et al. ~\cite{tuma2018threat} conducted a systematic literature review of 26 methodologies, evaluating their applicability, outcomes, and accessibility in the context of cyber threat analysis.


Recent research has broadened the landscape through NLP- and LLM-based security applications to automate and enhance CTI extraction, analysis, and downstream security tasks. Zhou et al.~\cite{zhou2022cti} combined BERT-based deep learning and NLP preprocessing for CTI analysis to automate IOC and TTP extraction and improve speed and accuracy over manual approaches. Studies employ transformer-based models such as BERT, RoBERTa, and domain-specific variants like SecureBERT that leverage contextual embeddings for higher precision and recall~\cite{aghaei2022securebert}. These models outperform classical ML approaches by capturing semantic relationships between technical terms and adversarial behaviors. LLM-centric literature studies, such as Xu et al.~\cite{xu2024large} and Zhang et al.~\cite{zhang2025llms}, have surveyed studies on LLM applications in cybersecurity, including vulnerability detection, malware analysis, and intrusion detection. These reviews highlight the expanding role of generative models in automating information extraction processes or improving their performance, but do not specifically explore TTP extraction. While Büchel et al. ~\cite{buchel2025sok} provided a SoK analysis of TTP extraction methods, their study focused primarily on benchmarking model performance across named-entity recognition, classification, and generative approaches. It offered a limited analysis of broader research trends, data source diversity, and methodological evolution. Thus, the study lacks a holistic synthesis of the field.

In summary, although existing studies have advanced understanding of CTI-centric reviews, sharing, and representation, to our knowledge, none of the studies position TTP extraction as the primary unit of analysis. Existing surveys treat TTPs as one extraction objective among many, without systematically analyzing the diversity of TTP-specific outputs, task formulations, evaluation rigor, or reproducibility practices. By shifting from a broad CTI-level synthesis to a dedicated, evaluation-aware review of ATT\&CK-aligned TTP extraction, our work fills a critical gap in the literature and offers a clearer methodological structure, identifies  research opportunities related to ontology ambiguity, data scarcity, model generalization, and scalable ATT\&CK-aligned solutions.

\section{Methodology}\label{4}
For this study, we have followed the guidelines by Kitchenham et al.~\cite{kitchenham2009systematic}, which provide methodological guidance for conducting systematic literature reviews and mapping of studies in software engineering. In this section, we discuss the steps of our methodology: i) Selection of Databases, ii) Search of Literature, iii) Selection of relevant studies, and iv) Analyzing the Studies.


\textbf{Selection of Databases:} To find relevant studies, we selected five widely recognized scholarly databases known for publishing high-quality research in computer science and related disciplines. Selected databases are IEEE Xplore~\cite{ieee}, ACM Digital Library~\cite{acm}, ScienceDirect~\cite{sciencedirect}, SpringerLink~\cite{springer} and Association for Computational Linguistics (ACL)~\cite{ACL}. We chose these databases based on their reputability, relevance to the domains of cybersecurity, machine learning, and natural language processing, and their large coverage of peer-reviewed publications. Moreover, in the field of computer science research, these scholarly databases are well known for conducting literature reviews~\cite{kuhrmann2017pragmatic}. Additionally, these databases index a wide range of high-impact conferences and journals, ensuring comprehensive coverage of relevant scholarly work.

\textbf{Search of Literature:}\label{search} 
We constructed a set of search strings by combining keywords and domain-specific terminology. The search strings were: (1) MITRE AND ATT\&CK, (2) MITRE OR ATT\&CK, and (3) Tactics AND Techniques AND Procedures. Since the goal of our search is to find specifically about the TTPs extraction from text, we specifically used only these aforementioned strings because any study performing TTP extraction would mention either ATT\&CK or TTPs at least once in the corresponding paper. We scoped our search strategy to MITRE ATT\&CK because it is a widely adopted and systematically maintained framework for modeling adversarial Tactics, Techniques, and Procedures (TTPs). ATT\&CK offers standardized, fine-grained technique definitions with stable identifiers, enabling consistent comparison across automated TTP extraction studies. We used these search strings in the selected scholarly databases mentioned in \textit{Selection of Databases}.  After initial search and duplicate removal, we found 3219 publications. The total number of papers obtained from each database is presented in Table~\ref{tab:search_result}. We performed a search for studies from 2015 to June 2025.

\begin{table}[ht]
\footnotesize
\caption{Search Results of Scholar Databases}
\label{tab:search_result}
\begin{tabular}{ll}
\toprule
Scholar Database    & Count \\ 
\midrule
IEEE                & 1228  \\ 
ACM Digital Library & 402   \\ 
ScienceDirect       & 539   \\ 
Springer Link       & 1016  \\ 
ACL                 & 34    \\ 
\bottomrule
\end{tabular}
\end{table}


 \textbf{Selection of Relevant Studies}
The initial search retrieved publications that were not explicitly focused on TTPs extraction. Hence, to maintain the studies' relevance to the intended scope and systematically exclude unrelated studies, we defined and applied inclusion and exclusion criteria. Filtration criteria are given as follows. Inclusion Criteria: (1) Publications must be available for downloading or reading online, (2) The title, keywords, abstract, and introduction of the paper should
explicitly indicate the publication’s relevance to TTPs extraction, (3) the publication must be written in English, and (4) the study should propose a novel approach or methodology for TTP extraction. Exclusion Criteria:  (1) Publications that are not peer-reviewed, such as keynote abstracts, call for papers, blogs, white papers
and presentations, (2) Duplicates of previously identified papers, (3) Published before the year 2015, and (4) Not related to TTPs extraction. The first two authors independently conducted the initial screening and filtering process to mitigate individual bias. All retrieved publications were manually assessed following the predefined inclusion and exclusion criteria described earlier. Following the filtration, 103 primary studies remained. To further enhance the completeness and coverage of the studies, we subsequently applied the established forward and backward snowballing technique ~\cite{wohlin2012experimentation} to the identified 103 articles. Forward snowballing systematically identified publications citing our initially selected studies, while backward snowballing examined the reference lists of these studies to identify cited publications. This iterative snowballing yielded an additional 31 relevant papers, bringing the corpus to 134 studies for further evaluation. The first two authors then independently conducted a comprehensive relevance assessment of all 134 papers to identify studies specifically addressing TTP  extraction methodologies. This assessment followed a systematic three-tier screening protocol: (1) initial screening of titles for topical relevance, (2) abstract review to evaluate methodological alignment with TTP extraction, and (3) full-text examination for articles where relevance could not be definitively determined from title and abstract alone. Through this multi-stage screening process, we excluded 54 publications that did not directly address TTP extraction or mining methodologies, resulting in 80 studies advancing to the final analysis phase. To validate the reliability and consistency of our independent selection process, we calculated Cohen's kappa coefficient, a widely used statistical measure of inter-rater agreement~\cite{cohen1960coefficient}. The obtained Cohen's kappa score of \texttt{0.86} indicates strong agreement between reviewers and demonstrates robust inter-rater reliability. Following these steps, we finally selected 80 relevant scholarly articles for further analysis for this study. 

 \noindent\textbf{Analyzing the Studies}
Following the stage of selection of studies, the first two authors independently conducted an in-depth systematic review of all 80 studies. During this phase, we systematically extracted and recorded the following information from each study: (a) TTP extraction purpose: the goal and specific security information extracted from textual sources and the insights derived from subsequent analysis, (b) Data source: the origin, nature, and characteristics of the employed datasets in the
studies, (c) Dataset Collection and Annotation: the process of data collection and annotation, (d) Methodology: the technical approaches, algorithmic frameworks, and procedural steps leveraged by the authors in their methodology, (e) Models and Metrics: the machine learning, NLP models or any other models used and the metrics applied for performance evaluation; (f) Limitations: the limitations of the study discussed by the authors. (g) Replicability and Availability: the assessment of reproducibility through verification of publicly available replication packages, code repositories, datasets, and implementation documentation for future studies.  
After completing the full-text review and data extraction, the first two authors applied a widely adopted qualitative open coding approach~\cite{richards2018practical} to synthesize and categorize the collected information. Since coding involves subjective judgment, we resolved disagreements using a card-sorting method~\cite{spencer2009card,tamanna2025your} combined with iterative discussion and refinement. Through iterative discussion and collaborative refinement, we reached an agreement on the following classifications: (a) 5 categories of TTP extraction purposes that address different threat intelligence goals
(Section 6); (b) 9 categories of data sources used for TTP extraction (Section 7); (c) 10 categories for data collection and preprocessing that represent different technical methods in the field (Section 8), (d) 6 categories of dataset annotation and construction for experiment (Section 9), (e) 6 categories of applied methodologies (Section 10). The details of the findings for each category are discussed in the subsequent sections.

\section{Summary of the Identified Papers}\label{5}
In this section, we present the results of the study selection process and introduce the studies. 
\textbf{Selected Studies}~\label{5.1}: 
As presented in Table~\ref{tab:search_result}, the number of publications retrieved from scholarly databases yields a total of 3,219 studies. The retrieved records were exported in CSV or BibTeX format. After removing duplicate entries, we applied the other inclusion and exclusion criteria as discussed earlier. This filtering process resulted in 80 publications that met all eligibility criteria for our study. The selected publications are listed in alphabetical order in Table~\ref{tab:ListofStudy} and are referenced throughout the paper using the notation P\#, such as P1 for the article \textit{A BERT-Based Framework for Automated Extraction of Behavioral Indicators of Compromise from Security Incident Reports}.

\noindent\textbf{Overview of Selected Study}~\label{5.2}: In this section, we provide an overview of the primary studies included in this systematic review. As shown in  Table~\ref{tab:ListofStudy}, the final set of 80 studies (P1–P80) is included in this study. The selected studies collectively investigated automated extraction of cyber threat intelligence (CTI) from textual and semi-structured sources, with a particular emphasis on adversarial TTPs. As illustrated by the titles, the corpus spans a range of methodological approaches, including traditional machine learning, deep learning, knowledge graphs, and, more recently, large language models and retrieval-augmented generation. This table serves as the foundation for subsequent analyses, in which we systematically compare the studies along dimensions such as extraction objectives, data sources, modeling techniques, and evaluation practices.

\begin{footnotesize}

\begin{longtable}{lp{130mm}}
\caption{List of Publications Selected for Systematic Literature Review}
\label{tab:ListofStudy}\\
\toprule
\textbf{Id} & \textbf{Title} \\
\midrule

P1 & A BERT-Based Framework for Automated Extraction of Behavioral Indicators of Compromise from Security Incident Reports~\cite{bekhouche2023bert} \\

P2 & A machine learning framework for investigating data breaches based on semantic analysis of adversary’s attack patterns in threat intelligence repositories~\cite{noor2019machineA} \\

P3 & A machine learning-based FinTech cyber threat attribution framework using high-level indicators of compromise~\cite{noor2019machineB} \\

P4 & A Novel Enhanced Naïve Bayes Posterior Probability (ENBPP) Using Machine Learning: Cyber Threat Analysis~\cite{sentuna2021novel} \\

P5 & A Pretrained Language Model for Cyber Threat Intelligence~\cite{park2023pretrained} \\

P6 & Actionable Cyber Threat Intelligence Using Knowledge Graphs and Large Language Models~\cite{fieblinger2024actionable} \\

P7 & Advancing TTP Analysis: Harnessing the Power of Large Language Models with Retrieval Augmented Generation~\cite{fayyazi2024advancing} \\

P8 & AECR: Automatic attack technique intelligence extraction based on fine-tuned large language model~\cite{chen2025aecr} \\

P9 & AEKG4APT: An AI-Enhanced Knowledge Graph for Advanced Persistent Threats with Large Language Model Analysis~\cite{zhou2025aekg4apt} \\

P10 & ALERT: A Framework for Efficient Extraction of Attack Techniques from Cyber Threat Intelligence Reports Using Active Learning~\cite{rahman2024alert} \\

P11 & AnnoCTR: A Dataset for Detecting and Linking Entities, Tactics, and Techniques in Cyber Threat Reports~\cite{lange2024annoctr} \\

P12 & APTKG: Constructing Threat Intelligence Knowledge Graph from Open-Source APT Reports Based on Deep Learning~\cite{sun2022aptkg} \\

P13 & Attack Behavior Extraction Based on Heterogeneous Cyberthreat Intelligence and Graph Convolutional Networks~\cite{tang2023attack} \\

P14 & Attack Behavior Extraction Based on Heterogeneous Threat Intelligence Graphs and Data Augmentation~\cite{li2024attack} \\

P15 & Attack Tactic Labeling for Cyber Threat Hunting~\cite{lin2022attack} \\

P16 & AttacKG: Constructing Technique Knowledge Graph from Cyber Threat Intelligence Reports~\cite{li2022attackg} \\

P17 & AttacKG+: Boosting attack graph construction with Large Language Models~\cite{zhang2025attackg+} \\

P18 & Automated Attacker Behaviour Classification Using Threat Intelligence Insights~\cite{crochelet2023automated} \\

P19 & Automated discovery and mapping ATT\&CK tactics and techniques for unstructured cyber threat intelligence~\cite{li2024automated} \\

P20 & Automated Emerging Cyber Threat Identification and Profiling Based on Natural Language Processing~\cite{marinho2023automated} \\

P21 & Automated Threat Report Classification over Multi-Source Data~\cite{ayoade2018automated} \\

P22 & Automatic Mapping of Unstructured Cyber Threat Intelligence: An Experimental Study: (Practical Experience Report)~\cite{orbinato2022automatic} \\

P23 & Building Cybersecurity Ontology for Understanding and Reasoning Adversary Tactics and Techniques~\cite{huang2022building} \\

P24 & ChronoCTI: Mining Knowledge Graph of Temporal Relations Among Cyberattack Actions~\cite{rahman2024chronocti} \\

P25 & Closing the Gap with APTs Through Semantic Clusters and Automated Cybergames~\cite{gianvecchio2019closing} \\

P26 & Clustering APT Groups Through Cyber Threat Intelligence by Weighted Similarity Measurement~\cite{chen2024clustering} \\

P27 & Construction of TTPS From APT Reports Using Bert~\cite{zongxun2021construction} \\

P28 & Crimson: Empowering Strategic Reasoning in Cybersecurity through Large Language Models~\cite{jin2024crimson} \\

P29 & CTIMiner: Cyber Threat Intelligence Mining Using Adaptive Multi-task Adversarial Active Learning~\cite{ma2025ctiminer} \\

P30 & Data Collection and Exploratory Analysis for Cyber Threat Intelligence Machine Learning Processes~\cite{wolf2022data} \\

P31 & Dataset of APT Persistence Techniques on Windows Platforms Mapped to the MITRE ATT\&CK Framework~\cite{rahal2025dataset} \\

P32 & DECEPT-CTI: A Framework for Enhancing Cyber Deception Strategies through NLP-based Extraction of CTI from Unstructured Reports~\cite{sayari2024decept} \\

P33 & DEEPCAPA: Identifying Malicious Capabilities in Windows Malware~\cite{vasan2024deepcapa} \\

P34 & Discovering attacker profiles using process mining and the MITRE ATT\&CK taxonomy~\cite{rodriguez2023discovering} \\

P35 & Enhanced small-scale APT knowledge graph embedding via spatio-temporal attribute reasoning and adversarial negative sampling~\cite{xie2025enhanced} \\

P36 & Enhancements to Threat, Vulnerability, and Mitigation Knowledge for Cyber Analytics, Hunting, and Simulations~\cite{hemberg2024enhancements} \\

P37 & Entity and relation extractions for threat intelligence knowledge graphs~\cite{mouiche2025entity} \\

P38 & Evaluating Text Augmentation for Boosting the Automatic Mapping of Vulnerability Information to Adversary Techniques~\cite{gionanidis2022evaluating} \\

P39 & Explainable cyber threat behavior identification based on self-adversarial topic generation~\cite{ge2023explainable} \\

P40 & Extracting Actionable Cyber Threat Intelligence from Twitter Stream~\cite{purba2023extracting} \\

P41 & Extraction of Threat Actions from Threat-related Articles using Multi-Label Machine Learning Classification Method~\cite{li2019extraction} \\

P42 & From Data to Action: CTI Analysis and ATT\&CK Technique Correlation~\cite{nguyen2024data} \\

P43 & GuardLink: Dynamic Linking of CVE to MITRE ATT\&CK Techniques using Machine Learning~\cite{el2024guardlink} \\

P44 & Identifying ATT\&CK Tactics in Android Malware Control Flow Graph Through Graph Representation Learning and Interpretability~\cite{fairbanks2021identifying} \\

P45 & Identifying Tactics of Advanced Persistent Threats with Limited Attack Traces~\cite{akbar2021identifying} \\

P46 & Improving Automated Labeling for ATT\&CK Tactics in Malware Threat Reports~\cite{domschot2024improving} \\

P47 & KnowCTI: Knowledge-based cyber threat intelligence entity and relation extraction~\cite{wang2024knowcti} \\

P48 & Labeling NIDS Rules with MITRE ATT\&CK Techniques Using ChatGPT~\cite{daniel2023labeling} \\

P49 & Leveraging BERT's Power to Classify TTP from Unstructured Text~\cite{alves2022leveraging} \\

P50 & Linking CVE’s to MITRE ATT\&CK Techniques~\cite{kuppa2021linking} \\

P51 & LLM-TIKG: Threat intelligence knowledge graph construction utilizing large language model~\cite{hu2024llm} \\

P52 & Looking Beyond IoCs: Automatically Extracting Attack Patterns from External CTI~\cite{alam2023looking} \\

P53 & Machine Learning Based Approach for the Automated Mapping of Discovered Vulnerabilities to Adversial Tactics~\cite{lakhdhar2021machine} \\

P54 & Malware2ATT\&CK: A sophisticated model for mapping malware to ATT\&CK techniques~\cite{sun2024malware2att} \\

P55 & MITREtrieval: Retrieving MITRE Techniques From Unstructured Threat Reports by Fusion of Deep Learning and Ontology~\cite{huang2024mitretrieval} \\

P56 & Modeling for Identifying Attack Techniques Based on Semantic Vulnerability Analysis~\cite{kim2024modeling} \\

P57 & Multi-label Classification of Cybersecurity Text with Distant Supervision~\cite{ishii2022multi} \\

P58 & Noise Contrastive Estimation-based Matching Framework for Low-Resource Security Attack Pattern Recognition~\cite{nguyen2024noise} \\

P59 & Not The End of Story: An Evaluation of ChatGPT-Driven Vulnerability Description Mappings~\cite{liu2023not} \\

P60 & Prompt Chaining-Assisted Malware Detection: A Hybrid Approach Utilizing Fine-Tuned LLMs and Domain Knowledge-Enriched Cybersecurity Knowledge Graphs~\cite{kumar2024prompt} \\

P61 & RAF-AG: Report analysis framework for attack path generation~\cite{mai2025raf} \\

P62 & Reproducing ATT\&CK Techniques and Lifecycles to Train Machine Learning Classifier~\cite{yudha2025reproducing} \\

P63 & Safeguarding Infrastructure from Cyber Threats with NLP-Based Information Retrieval~\cite{salley2023safeguarding} \\

P64 & Semantic Ranking for Automated Adversarial Technique Annotation in Security Text~\cite{kumarasinghe2024semantic} \\

P65 & STIXnet: A Novel and Modular Solution for Extracting All STIX Objects in CTI Reports~\cite{marchiori2023stixnet} \\

P66 & Threat Behavior Textual Search by Attention Graph Isomorphism~\cite{bae2024threat} \\

P67 & Threat intelligence ATT\&CK extraction based on the attention transformer hierarchical recurrent neural network~\cite{liu2022threat} \\

P68 & Threat Miner - A Text Analysis Engine for Threat Identification Using Dark Web Data~\cite{deguara2022threat} \\

P69 & ThreatKG: An AI-Powered System for Automated Open-Source Cyber Threat Intelligence Gathering and Management~\cite{gao2023threatkg} \\

P70 & TIM: threat context-enhanced TTP intelligence mining on unstructured threat data~\cite{you2022tim} \\

P71 & Towards Effective Identification of Attack Techniques in Cyber Threat Intelligence Reports using LLM~\cite{cuong2025towards} \\

P72 & TREC: APT Tactic / Technique Recognition via Few-Shot Provenance Subgraph Learning~\cite{lv2024trec} \\

P73 & TTPatternMiner: Automated Learning and Characterization of Attack Pattern from Malicious Cyber Campaign~\cite{ullah2024ttpatternminer} \\

P74 & TTPDrill: Automatic and Accurate Extraction of Threat Actions from Unstructured Text of CTI Sources~\cite{husari2017ttpdrill} \\

P75 & TTPHunter: Automated Extraction of Actionable Intelligence as TTPs from Narrative Threat Reports~\cite{rani2023ttphunter} \\

P76 & TTPMapper: Accurate Mapping of TTPs from Unstructured CTI Reports~\cite{ali2024ttpmapper} \\

P77 & TTPXHunter: Actionable Threat Intelligence Extraction as TTPs from Finished Cyber Threat Reports~\cite{rani2024ttpxhunter} \\

P78 & Uncovering Hidden Threats: Automated, Machine Learning-based Discovery \& Extraction of Cyber Threat Intelligence from Online Sources~\cite{ellinitakis2024uncovering} \\

P79 & Using Natural Language Processing Tools to Infer Adversary Techniques and Tactics Under the Mitre ATT\&CK Framework~\cite{gabrys2024using} \\

P80 & V2TSA: Analysis of Vulnerability to Attack Techniques Using a Semantic Approach~\cite{kim2024v2tsa} \\

\bottomrule

\end{longtable}

\end{footnotesize}

\section{Paper Categories}\label{6}
\label{papercategories}
In this section we answer {\textit{RQ1: What are the primary objectives of TTP-related information extraction?}}
To systematically examine the goal of prior TTP extraction research, we reviewed and classified all selected studies according to their primary analytical objectives in terms of the specific TTP-related information produced and the corresponding modeling formulation.

For each paper, we analyzed the stated extraction goals, methodological design and the structure of the reported outputs. This process enabled us to determine both what level of TTP abstraction was targeted and how the task was operationalized computationally. Based on the analysis, we organized the studies into five categories: \textit{Tactic Classification, Technique Classification, Technique Searching, IOC Extraction \& TTP, and Knowledge Graph (KG) Construction \& TTP}. Since many studies pursued multiple objectives (e.g., performing technique classification followed by knowledge graph construction), we adopted a multi-label categorization scheme in which a study may belong to multiple categories. Applying this approach reflects the methodological overlap and complexity of the literature. The distribution of studies across the five categories is summarized in Table~\ref{tab:paper_categorization} and discussed in detail in the following subsections.

\begin{table}[H]
\centering
\footnotesize
\caption{Categorization of Selected Papers by Analysis Objective}
\label{tab:paper_categorization}
\begin{tabular}{p{4cm} p{8cm}}
\toprule
\textbf{Paper Category} & \textbf{Paper IDs} \\
\midrule
Tactic Classification &
P7, P15, P20, P45, P46, P53 \\

Technique Classification &
P5, P8, P10, P14, P18, P19, P21, P22, P25, P27, P28, P31, P33, P34, P38, P39, P41--P44, 
P48--P50, P54--P59, P62, P67, P70--P72, P74--P77, P79, P80 \\

Technique Searching &
P2, P3, P4, P64, P66 \\

IOC Extraction \& TTP &
P1, P26, P32, P40, P73, P78 \\

Knowledge Graph (KG)  Construction \& TTP &
P6, P9, P11--P14, P16, P17, P23, P24, P29, P30, P35--P37, P47, P51, P52, 
P60, P61, P63, P65, P68, P69 \\
\bottomrule

\end{tabular}
\end{table}

\subsection{Tactic Classification (n=6)}
This category comprises studies that focus on the automated classification of CTI text or related artifacts into adversarial tactics defined in the MITRE ATT\&CK framework. We identified six studies proposing automated approaches for mapping inputs to one or more ATT\&CK tactics. Based on the type of input artifact and analytical objective, these studies fall into two subcategories.


The first group includes studies (P20, P46, P53) that classify textual artifacts into ATT\&CK tactics. These studies classify sentences drawn from sources such as social media posts, CTI reports, and National Vulnerability Database (NVD)~\cite{nist_nvd} Common Vulnerabilities and Exposures (CVE) descriptions into corresponding tactics. The studies primarily operate on narrative or descriptive text and emphasize tactic-level abstraction over fine-grained technique identification.
The second group (P15, P45, P7) focuses on classifying non-textual or structured operational artifacts into \attack tactics. P15 maps SNORT rules used in network intrusion detection systems to tactics using machine learning, while P45 infers tactics from system provenance graphs using machine learning. In addition, P7 evaluates the effectiveness of different large language model architectures by comparing encoder-only and decoder-only models for tactic classification.

\subsection{Technique Classification (n=39)}
This category focuses on the automated classification of CTIs' text into relevant adversarial Techniques. We identified 39 studies that proposed automated approaches for mapping such textual artifacts to MITRE ATT\&CK techniques. Based on the primary input data source, we identified five subcategories of papers as shown in Table~\ref{tab:ttps extraction paper types}. 

\begin{table}[ht]
\footnotesize
\caption{TTPs Extraction Paper types}
\label{tab:ttps extraction paper types}
\begin{tabular}{lp{8cm}}
\toprule
\textbf{TTPs Extraction Types} & \textbf{Papers}                                                                                                      \\ \midrule
CTI Reports to Techniques & P5, P8, P10, P14, P21-22, P27-28, P39, P41-42, P49, P55, P57-58, P62, P67, P70-72, P74-77 \\
CVE Entries to Techniques      & P38, P43, P50, P56, P59, P80      \\
System logs to Techniques      & P18, P25, P31, P33-34 \\
NIDS Rules to Techniques       & P48, P79 \\
Malware Source Code to Techniques       & P44, P54 \\ \bottomrule
\end{tabular}
\end{table}

\begin{enumerate}[label=(\alph*)]
    \item \textbf{CTI Reports (n=24): } This subcategory of papers classifies the sentences in CTI reports to one or more MITRE \attack techniques. 
    \item \textbf{CVE Entries (n=6): } This subcategory of papers classifies the sentences in vulnerability descriptions to one or more MITRE \attack techniques. 
    \item \textbf{System Logs (n=5): } This subcategory of papers classifies the system and network events, API calls to one or more MITRE \attack techniques. 
     
    \item \textbf{NIDS Rules (n=2): } This subcategory of papers classifies the YARA and SNORT rules to one or more MITRE \attack techniques. 
    \item \textbf{Malware Source Code (n=2): } This subcategory of papers classifies the modules of malware source codes to one or more MITRE \attack techniques.
\end{enumerate}

As shown in Table~\ref{tab:ttps extraction paper types}, the majority of technique classification studies relied on unstructured or semi-structured CTI reports, accounting for 24 papers (30\% of the full corpus). This predominance reflects the central role of narrative threat reports in contemporary CTI workflows and their suitability for NLP-based analysis. A smaller subset of studies maps vulnerability descriptions from CVE entries to ATT\&CK techniques (7.5\%), aiming to bridge vulnerability intelligence with adversarial behavior modeling. Fewer studies operate on system or network system logs (6.25\%), positioning their approaches closer to operational detection and monitoring pipelines.

Further, a smaller number of studies mapped network intrusion detection system (NIDS) rules to ATT\&CK techniques or inferred techniques directly from malware source code (2.5\% each), indicating comparatively lower research attention to code-level and detection-rule–based inputs. Overall, this distribution highlights a strong emphasis on CTI report–driven extraction, with substantially less coverage of operational artifacts such as logs, source code, and detection rules. This imbalance suggests clear opportunities for future research to better support real-time detection, incident response, and operational security workflows.


\subsection{Technique Searching (n=5)}
CTI text often contains step-by-step descriptions of attack activities, including what actions were performed, who conducted them, their motivations, and how the attacks were executed. However, such text is typically highly technical, heterogeneous in style, and tailored to diverse audiences, ranging from operational details for security operations centers (SOCs) to high-level strategic summaries for executive stakeholders. As a result, identifying relevant ATT\&CK techniques within large volumes of unstructured CTI text poses a great challenge for cybersecurity practitioners to find the techniques used by adversaries.

We identified five studies that address this problem by proposing automated approaches for searching, rather than classifying, ATT\&CK techniques in CTI text. Based on their retrieval strategy, these studies fall into two subcategories. The first group of studies (P64 and P66) implements custom search engines designed to retrieve \attack techniques from CTI reports using rule-based or keyword-driven matching. The second group, consisting of (P2, P3, P4), adopts semantic search approaches that leverage distributed representations to retrieve techniques based on contextual similarity. These semantic retrieval methods are often employed as a precursor to downstream analysis tasks, such as cyber threat actor attribution. Overall, the limited number of studies in this category indicates that technique searching remains relatively underexplored compared to classification-based approaches.


\subsection{Indicators of Compromise Extraction \& TTP (n=6)} 
CTI frequently includes fine-grained signals of security incidents or malicious behavior, commonly known as Indicators of Compromise (IoCs). Typical IoCs include IP addresses, domain names, file hashes (e.g., MD5, SHA1, SHA256), email addresses, registry keys, and other forensic artifacts. These indicators are frequently embedded in unstructured sources such as threat reports, security blogs, social media posts, and dark web forums. Automated extraction of IoCs from CTI text is therefore a critical capability for enabling timely threat detection, enrichment, and response. We grouped the six IoC extraction studies into three subcategories based on the semantic granularity of extracted indicators and the underlying CTI sources.

 The first group of studies (P1, P73, P26) focuses on behavior-oriented indicators that closely align with adversarial actions and ATT\&CK concepts. P1 extracts tactics, techniques, sub-techniques, adversary groups, and tools from CTI reports, while P73 targets higher-level behavioral indicators such as actions, intents, configurations, tools, and potential actions. Similarly, P26 extracts techniques and software alongside contextual attributes such as target nations and affected industries. Collectively, these studies emphasize semantic understanding of adversary behavior rather than low-level forensic artifacts. In contrast, the second group (P40, P78) prioritizes the extraction of concrete, low-level indicators commonly used for detection and blocking. P40 analyzes social media content (X posts) to extract hashtags, URLs, embedded objects, and related indicators, whereas P78 examines dark web forums to extract infrastructure-centric IoCs, including malware, tools, threat actors, identities, and locations. These approaches focus on producing actionable artifacts that can be directly operationalized in security tools. Finally, the last one (P32) adopts a comprehensive extraction strategy, aiming to cover a broad range of IoC types within a single pipeline. This study extracts domains, IP addresses, malware, hashes, CVE identifiers, threat actors, temporal markers, operating systems, and tools, and explicitly targets downstream decision-making tasks, such as selecting optimal cyber deception strategies. Unlike the other studies, this approach emphasizes breadth of coverage and integrative analysis over specialization in a single indicator class.

Across all groups, we observe that the extracted indicators largely conform to STIX 2.1 domain object schema, suggesting convergence toward standardized CTI representations. However, most studies focus on select subsets of IoCs, particularly techniques, software, and threat actors, rather than comprehensive schema coverage. This uneven emphasis indicates that while IoC extraction is well-studied for high-value indicators, holistic extraction across the full STIX remains unexplored.


\subsection{Knowledge Graph Construction \& TTP (n=24)}
This category includes studies that construct a Knowledge graph (KG) from CTI text. Knowledge graph (KG) construction enables the transformation of unstructured CTI text into structured representations that explicitly encode entities, relations, and contextual dependencies defined in standardized threat intelligence schemas such as STIX and MITRE ATT\&CK. By organizing extracted tactics, techniques, indicators, and actors into machine-interpretable graphs, these approaches support integration across heterogeneous CTI sources as well as downstream reasoning and analysis. We identified 24 such papers and grouped them into four subcategories based on the role the knowledge graph within the overall analysis pipeline

\noindent \textbf{Extraction of KG (n=20):} This subcategory of papers extracted KG from CTI-text, specifically entities and relations among the entities. We identify twenty (P6, P9, P12, P13, P14, P16, P17, P23, P24,P29, P30, P37, P51, P52, P60, P61, P63, P65, P68, P69) papers in this subcategory. Table~\ref{tab:kg_extraction_paper_types} shows the papers in this subcategory. We observe that most papers extracted KGs from CTI reports, except for a few cases, such as system logs and memory dumps. We observe that most of the papers extracted entities and relations defined in the STIX version 2 ontologies. However, none of the studies extracted the full spectrum of defined entities and relations from STIX. The most commonly extracted entities are technique, malware, indicators, actors, and vulnerability. The most commonly extracted relations are targets, indicates, use, and mitigates. Apart from STIX, we also observe several special types of KGs, such as temporal relations, narrative orders, and dependency. Finally, we also observe several cases where the types of extracted entities and relations were not specified.
    
\noindent \textbf{KG Completion (n=2)}: This subcategory of papers proposes approaches for automatically predicting the missing links among the entities in an existing KG. We identify two (P35, P47) papers in this subcategory. Study P35 proposed a KG embedding approach for completing missing links between nodes in existing KGs extracted from MITRE \attack Flow~\cite{mitre_engenuity_attack_flow}. Research article P47 proposed a KG embedding approach for completing missing links between nodes in existing KGs extracted from OpenCVE~\cite{opencve_platform}, CAPEC~\cite{mitre_capec}, and MalVuln~\cite{malvul_db}. 
    
\noindent \textbf{KG Linking (n=1)}: This subcategory of papers proposes approaches for unifying multiple KGs generated based on different ontologies. We identify one (P36) paper in this subcategory that unified several well-known ontologies into a single knowledge graph system, BRON, which integrates MITRE \attack, CAPEC~\cite{mitre_capec}, and CVE~\cite{opencve_platform}. They further linked graph entities to ExploitDb~\cite{offsec_exploit_db} and MITRE CAR~\cite{mitre_car} for entity disambiguation. 
    
\noindent \textbf{KG Dataset (n=1)}: This subcategory of papers constructs a dataset for training a machine learning model for KG extraction from CTI-text. We identify one (P11) paper that proposed a dataset of annotated CTI reports comprising the following entities: organization, location, sector, timex, code snippets, group, malware, tool, concept, tactic, technique. They annotated the connections between entities: temporal (date and time) and disambiguation (Wikipedia and \attack).


The analysis indicates that most TTP extraction research concentrates on fine‑grained technique identification, reflecting the operational need to map CTI text to specific MITRE ATT\&CK techniques for detection and automation. In contrast, fewer studies address higher‑level tactic classification, technique search, or IOC–TTP correlation. Knowledge graph construction has recently gained traction as a way to model relationships among threat entities and attack behaviors, but it remains less explored than technique classification.


\begin{table}[htbp]
\footnotesize
\caption{Extraction of KG Papers}
\setlength{\tabcolsep}{1.5pt}
\label{tab:kg_extraction_paper_types}
\begin{tabular}{lp{28mm}p{5cm}p{5cm}}
\toprule
\textbf{Paper} &
  \textbf{Source} &
  \textbf{Entities} &
  \textbf{Relations} \\ \midrule
P6 &
  CTI Reports &
  Malware Types, Applications, Operating Systems, Organizations, Persons, Times, Threat Actors, Locations, and Attack Patterns &
  isA, targets, uses, hasAuthor, hasAlias, indicates, discoveredIn, exploits, variantOf, and has \\
P9 &
  CTI Reports &
  attacker, attack pattern, campaign, course of action, identify, indicator, observed data, malware, tool, location, vulnerability &
  attributed to, related to, use, impersonate, target, located at, exploit, mitigate, indicate, based on, authored by, download \\
P12 &
  CTI Reports &
  group, attacker, technique, malware, tool, ip, network connection, file, key, vulnerability, course of action, victim, location &
  Uses, Indicates, Targets, Based-on, Communicates-with, Mitigates, Located-at, Has, and About \\

P13 & 
  CTI Reports &
  techniques, tactics, adversary, and tools &
  not specified \\
  
P14 &
  CTI Reports &
  not specified &
  not specified \\
P16 &
  CTI Reports &
  technique, actor, executable, file, network connections, registry, and others &
  dependency \\
P17 &
  CTI Reports &
  tactic/technique, attacker, malware, c2, campaign, attack &
  use, target, communicate with, execute, host \\
P23 &
  CTI Reports &
  tactice, technique, software, group, mitigation &
  hasTechnique supportsRemote, hasId, accomplishesTactic \\
P24 &
  CTI Reports &
  technique &
  before, concurrent, simultaneous \\
P29 &
  CTI Reports, hacker forum &
  not specified &
  not specified \\
P30 &
  Unknown &
  Tool, Indicator , Report , Malware , Course of Action , Attack Pattern, Vulnerability , Intrusion Set , Identity , Campaign , Threat Actor &
  To, Targets, Indicates, Uses, Applies, Mitigates, Related To, Attributed To, Part Of, Created By, Revoked By. Comprontised By \\
P37 &
  CTI Reports &
  HackOrg, OffAct, SamFile, SecTeam, Tool, Time, Purp, Area, Idus, Org, Way, Exp, and Features, application, cveID, edition, file, function, hardware, method, OS, parameter, programming language, relevant term, update, vendor, and version, malware, indicators of compromise, system, organization, and vulnerability &
  associatedWith:, discovers:, discoveredBy:, hasAttackTime:, hasCharacteristics:, locatedAt:, monitors:, monitoredBy:, uses, usedby, targets, targetedby \\
P51 &
  CTI Reports &
  malware, threat type, attacker, technique, tool, vulnerability, ip. Domain, url., file, hash &
  not specified \\
P52 &
  CTI Reports &
  Malware, Malware Type, Application, Operating System, Organization, Person, Time, Threat Actor, Location, and Attack Pattern. &
  isA, targets, uses, hasAuthor, hasAlias, indicates, discoveredIn, exploits, variantOf, and has \\
P60 &
Memory dump, network packet &
  campaigns, dataset, group, mitigation, software, tactic, technique &
  use tactic, attributed to, uses, mitigates, detects \\
P61 &
  CTI Reports &
  techniques &
  narrative order of techniques \\
P63 &
  ATT\&CK corpus &
  techniques &
  path \\
P65 &
  CTI Reports &
  attack patterns, campaign, identity, intrusion set, location, malware, tool &
  not specified \\
P68 &
  Hacker forum &
  not specified &
  not specified \\
P69 &
  CTI Reports &
  not specified &
  not specified \\ \bottomrule
\end{tabular}
\end{table}

\newpage
\section{Data Sources}\label{7}

In this section,  we answer \textit{RQ2: What sources have been used for data collection?} To systematically analyze how TTPs extraction studies obtain their input data, we analyzed the source, purpose, and type fields of all selected studies to identify their data origins for TTPs extraction. The sources were grouped into eight categories. If a study used multiple data types, it was included in all relevant categories to represent the diversity of data sources used in TTPs extraction research. The data source categories are presented in Table~\ref{tab:data_sources_grouped} and discussed in detail below. 

\textbf{Benchmark Datasets and Public Knowledge Base (n = 48)}: Benchmark datasets and public knowledge bases provide standardized and reproducible environments for developing and evaluating TTP extraction and detection models. Most used datasets and knowledge bases are MITRE ATT\&CK~\cite{mitre} (a structured knowledge base of tactics and techniques), CAPEC~\cite{mitre_capec} (common platform enumeration for specifying hardware and software), CWE~\cite{mitre_cwe} (common software security weaknesses), TRAM~\cite{mitre_ctid_tram} (a reputed dataset for sentences to TTPs mapping), where they provide labeled and expert-curated corpora for evaluating TTPs extraction models. These resources typically contain labels and well-defined schemas, support large-scale experimentation, comparative evaluation of model performance, and reproducibility across studies.

 \textbf{CTI Reports (n = 28)}: CTI reports are used for threat intelligence extraction and TTPs identification. These reports are often released by security vendors and research organizations, such as FireEye~\cite{fireeye_reports}, Kaspersky~\cite{kaspersky_apt_report}, and Symantec~\cite{symantec}, and provide narrative descriptions of threat campaigns, adversary behaviors, and associated indicators of compromise. Their rich contextual detail makes them particularly suitable for natural language processing–based extraction and classification tasks.

\textbf{Vulnerability Database (n = 8)}: Structured vulnerability data sources facilitate mapping known weaknesses and detection signatures to adversarial techniques. Commonly used sources include the National Vulnerability Database (NVD)~\cite{nist_nvd} and ENISA~\cite{enisa_euvd} vulnerability corpora. These datasets typically contain standardized vulnerability descriptions and metadata, enabling studies to link vulnerabilities to ATT\&CK techniques and bridge vulnerability intelligence with behavioral threat modeling.

\textbf{System and Network Logs (n = 9)}: System and network logs provide fine-grained, execution-level evidence of adversary activity and are often used to infer techniques through event correlation and behavioral analysis. Data sources in this category include Sysmon and EVTX logs, system provenance graphs, and honeypot datasets such as Cowrie~\cite{oosterhof_cowrie}. These logs capture host and network-level events that support the reconstruction of attack sequences and are commonly used in technique classification and detection-oriented studies.

 \textbf{Intrusion Detection System Rules (n = 4)}: Signature-based Intrusion Detection System (IDS) rulesets, including Wazuh IDS~\cite{wazuh_wazuh_ruleset} and SNORT~\cite{snort_vrt_rules}, serve as structured data representations of known attack patterns. These rule sets have been used to align detection signatures with ATT\&CK techniques, evaluate rule coverage, and construct labeled datasets that link detection logic to adversarial behavior. As such, they provide a direct connection between operational security tooling and TTP modeling.

\textbf{Threat Intelligence Platform (n=3)}: Threat Intelligence Platforms (TIPs) are centralized systems that aggregate, normalize, and enrich threat data from diverse sources, including open-source feeds, dark web content, and internal telemetry. Among the selected studies, AlienVault OTX~\cite {LevelBlue_OTX_2025} was identified as a commonly used platform. TIPs provide integrated views of threat data and are leveraged to support multi-source TTP extraction and correlation.

\textbf{Malware Repositories (n=3)}: Malware repositories provide access to large collections of known malware samples and associated behavioral data. Commonly used sources include VirusShare~\cite{virusshare_repo} and VirusTotal~\cite{virustotal_platform}, which offer static and dynamic analysis outputs. These datasets are used to extract malware behaviors and map them to TTPs, particularly for studying persistence, privilege escalation, and execution-related techniques.

 \textbf{Cybersecurity Books and Academic Papers (n=2)}: Cybersecurity textbooks and peer-reviewed academic publications serve as meta-sources for TTP extraction research. These sources are primarily used to synthesize formal definitions, derive annotation guidelines, and compare extraction paradigms. While less frequently used as primary input data, such documents provide authoritative context and conceptual grounding for TTP modeling and extraction.

\begin{table}[htbp]
\centering
\scriptsize
\setlength{\tabcolsep}{4pt}
\renewcommand{\arraystretch}{0.9}

\caption{Data sources used in TTP extraction studies}
\label{tab:data_sources_grouped}

\begin{tabular}{p{0.23\textwidth}p{0.42\textwidth}p{0.30\textwidth}p{0.05\textwidth}}
\toprule
\textbf{Source Category} & \textbf{Sources} & \textbf{Studies} & \textbf{Count} \\
\midrule

Benchmark Datasets and Public Knowledge Base &
MITRE ATT\&CK, CAPEC, CWE, TRAM, APTNotes, MalwareTextDB, CyberFed Model, STIG bundles, BRON dataset, DNRTI, STUCCO, CyNER, CREME v2 dataset, rcATT repository, Atomic Red Team &
P1, P2, P4, P5, P7, P9, P10, P12, P13, P14, P15, P16, P17, P19, P21, P22, P23, P24, P26, P27, P29, P30, P32, P35, P36, P37, P39, P41, P42, P46, P53, P55, P56, P58, P59, P60, P62, P63, P64, P65, P67, P68, P69, P71, P75, P76, P77, P80 &
48 \\

CTI Reports &
FireEye, CrowdStrike, Symantec, Kaspersky Securelist, Cisco Talos Intelligence, Microsoft Security Center, Intel471, Lab52, QuoIntelligence, DFIR Reports, Cyble, Fortinet, Trend Micro, Palo Alto Networks, McAfee, ESET Threat Reports, IBM X-Force, The Hacker News, ThreatPost, CitizenLab, KrebsOnSecurity, Security Affairs, WeLiveSecurity, BrightTalk, InfoSecurity, Proofpoint, Forcepoint, Novetta, ClearSkySec, ThreatConnect, Google TAG, AT\&T Cybersecurity, KnowB4 &
P1, P2, P3, P5, P6, P8, P11, P16, P17, P21, P24, P26, P32, P47, P49, P50, P51, P52, P55, P57, P61, P64, P66, P67, P69, P70, P73, P74 &
28 \\

Vulnerability Database &
National Vulnerability Database (NVD), ENISA Vulnerability Dataset, CVEList &
P5, P28, P36, P38, P43, P53, P59, P80 &
8 \\

System and Network Logs &
Cowrie Honeypot Logs, BRAWL Cyber Gameboard Logs, Windows System Logs, Memory Snapshots, PWNJUTSU Dataset, Atomic Red Team Execution Logs, CIC-IDS2017, CIC-MalMem-2022, CREME v2 Dataset &
P18, P25, P31, P33, P34, P45, P60, P62, P72 &
9 \\

Intrusion Detection System Rules &
Snort Rules, Wazuh IDS Ruleset, Suricata Rules &
P15, P48, P76, P79 &
4 \\

Threat Intelligence Platforms &
AlienVault OTX, 360TIP, ALPHA Threat Intelligence Platform, ThreatMiner &
P3, P9, P76 &
3 \\

Malware Repositories &
VirusTotal, VirusShare, Malpedia &
P9, P44, P54 &
3 \\

Social Media and Dark-Web Forums &
Twitter/X OSINT, BreachForums, CrimeBB dataset, Hacker forum dataset &
P20, P29, P40, P78 &
4 \\

Cybersecurity Books and Academic Papers &
CISSP Textbooks, USENIX Security Papers, Security Wiki &
P5, P41 &
2 \\

\bottomrule
\end{tabular}

\end{table}

The majority of studies rely primarily on benchmark datasets, public knowledge bases, and CTI reports, with the MITRE ATT\&CK framework among the most commonly used sources. In contrast, other data sources, such as vulnerability databases and system and network logs, are used less frequently, resulting in relatively limited diversity in the datasets employed across the literature.



\section{Data Collection and Preprocessing}\label{8}
In this section, we answer \textit{RQ3: How is the dataset collected and pre-processed?} We systematically analyzed the data collection strategies and preprocessing techniques employed across the selected studies. For data collection, we organized the findings into four categories, while preprocessing practices were grouped into six categories. These categories reflect recurring methodological patterns observed in the literature. The findings are presented along these key dimensions to highlight common practices, design trade-offs, and gaps in current approaches.

\subsection{Data Collection}
Our study identifies four categories of data collection approaches. The details of each category, along with the number of associated studies, are discussed below.

\textbf{Automated Web Crawling and Large-Scale Scraping (n=10):} Several studies (P21, P51, P52, P56, P69, P70, P73, P74, P78) employed automated web crawlers or scrapers (e.g., \textit{newspaper3k}\footnote{https://newspaper.readthedocs.io/en/latest/}) as the primary means of data collection. The crawlers and scrapers target open web sources such as threat intelligence blogs, vendor advisories, malware analysis reports, and cybersecurity news articles. The use of large-scale web crawlers enables high recall and enables the inclusion of diverse artifacts in adversarial campaigns. However, crawler-based data collection contains substantial noise, as scraped content often includes boilerplate text, advertisements, navigation elements, and duplicated articles syndicated across multiple sources. As a result, while web crawling facilitates scalability, it requires post-collection filtering. To mitigate noise, P74 integrates an SVM-based classifier to filter and retain only CTI-relevant articles, whereas P53 focuses on removing duplicates to remove redundant content.

\textbf{Custom Search Engine and API Based Collection (n=2):} Two studies (P3, P20) adopted a retrieval strategy using custom search engines and platform APIs. P3 employed a custom Google search engine restricted to cybersecurity-relevant domains, while P20 collects social media data using Logstash X API\footnote{https://www.elastic.co/docs/reference/logstash/plugins/plugins-inputs-twitter} based API pipelines. This approach prioritizes topical precision and temporal freshness, enabling focused collection around explicit threat-related terms such as APT group names, malware aliases, or CVE identifiers. While more constrained in scale, API-driven collection offers tighter control over relevance and timeliness.

\textbf{Credibility-Oriented Data Collection (n=2):} Two studies (P1, P12) emphasized expert curation and source credibility as primary data selection criteria These studies emphasize two criteria: (i) credibility favoring institutional or vendor-authored threat reports (e.g., FireEye\cite{fireeye_reports}, CrowdStrike\cite{crowdstrike_reports}), and (ii) coverage breadth of broad types of APT attacks, and malware. Such credibility-oriented strategies reflect a design choice that favors analytical reliability and trustworthiness, particularly for studies focused on behavioral interpretation rather than large-scale automation.

\textbf{License-Constrained Collection(n=1):}
Only one study (P11) explicitly addresses licensing and legal reuse constraints, especially by restricting data collection to Creative Commons–licensed sources (e.g., CC-BY, CC-BY-SA 4.0). Although this constraint limits corpus size, it substantially improves reproducibility and long-term accessibility, a critical but often overlooked issue in CTI research, where proprietary data and link rot are common. This approach highlights an important direction for building sustainable and shareable CTI datasets.

\subsection{Data Preprocessing}
Our study identifies six categories of data collection approaches. The details of each category, along with the number of associated studies, are discussed below.

\textbf{Use of Parser Libraries and DOM-Level Extraction (n=4):} Parser libraries such as BeautifulSoup\footnote{https://beautiful-soup-4.readthedocs.io/en/latest/}, and Apache Tika\footnote{https://tika.apache.org} are prevalent in CTI text extraction pipelines (P8, P11, P57, P70). These tools enable Document Object Model (DOM) level parsing and hierarchical decomposition of web pages and documents, allowing studies to selectively extract relevant textual components such as paragraphs, headings, list items, and metadata fields while discarding non-informative elements (e.g., code snippets, navigation menus, or embedded scripts). Such structured extraction supports cleaner downstream text analysis by preserving semantic anchors and document structure.

\textbf{Conversion from PDF to Plain Text (n=3):} Studies (P1, P32, and P74) report a preprocessing stage that converts heterogeneous file formats(i.e., PDF, DOCX) into plain text. This normalization step removes layout-specific artifacts, including figures, tables, and formatting metadata, to produce a uniform textual representation. However, this conversion is particularly challenging in CTI contexts, as PDFs often contain structured IoC tables, multi-column layouts, or mixed character encodings. These characteristics can fragment sentences or disrupt token order, potentially degrading downstream extraction performance.

\textbf{Text Cleaning and Lexical Standardization (n=5):} Traditional text-cleaning operations, such as stop-word removal, punctuation filtering, case normalization, and lemmatization, are applied in several studies (P1, P19, P20, P22, P36). While these techniques reduce noise and vocabulary sparsity, they can inadvertently suppress domain-specific tokens that convey security semantics, such as file extensions, registry paths, hexadecimal strings, or command-line artifacts. This underscores a balancing challenge between linguistic normalization and retaining CTI-relevant signals.

\textbf{Sentence Segmentation and Tokenization (n=12):} Sentence segmentation and tokenization are widely adopted preprocessing steps, reported in twelve studies (P5, P6, P11, P14, P19, P20, P23, P32, P36, P42, P45, P49). These steps transform normalized text into linguistically coherent units, which are essential for dependency parsing, relation extraction, and transformer-based representation learning. Two studies (P11, P19) employ custom rule-based sentence segmenters customized to CTI-specific text, where conventional punctuation-based boundaries fail (e.g., “CVE-2021-40444” or “APT28/29”). Additionally, P6 explicitly considers model input constraints by enforcing token window limits compatible with transformer architectures (e.g., 4096 tokens).

\textbf{IOC-Specific Text Processing (n=4):} Several studies (P19, P23, P70, P77) extracted or redacted Indicator of Compromise (IoC) handling. Since IoCs (e.g., hashes, IP addresses, URLs) can be both semantically informative and potentially sensitive, studies adopt different handling strategies. Some replace IoCs with symbolic placeholders (e.g., “<IP>'', “<HASH>'') to reduce sparsity or mitigate leakage risks, while others retain and tokenize IoCs to enable structured extraction and classification. These divergent strategies reflect differing priorities between simplification and representational accuracy.

\textbf{Regular Expression, and Heuristic-Based Cleaning (n=3):} Three studies (P11, P32, and P67) used regular expressions and rule-based heuristics to detect and remove noisy patterns such as inline citations, code blocks, or figure captions. Although these approaches are labor-intensive and lack scalability, they offer high interpretability and precision, particularly when dealing with irregular or non-standardized CTI documents. Their continued use suggests that rule-based preprocessing remains effective in scenarios where robustness and transparency are prioritized.

Overall, dataset construction typically involves extracting sentences or documents from CTI reports and mapping them to MITRE ATT\&CK techniques or tactics, often followed by filtering and preprocessing steps such as sentence segmentation, entity replacement, and noise removal. Dataset preparation in TTP extraction is largely manual and annotation-intensive, with most studies using sentence-level datasets derived from CTI reports and ATT\&CK documentation.

\section{Dataset Annotation and Construction for Experiment}\label{9}
In this section, we answer \textit{RQ4: How is the dataset annotated and constructed for the experiment?} We examine how selected studies annotate CTI data and construct labeled datasets to support supervised and semi-supervised TTP extraction tasks. Our analysis focuses on annotation approaches, annotation tools and frameworks, quality assurance through inter-annotator agreement mechanisms, data augmentation techniques, ground truth and labeling sources, and dataset structuring and representation formats. Together, these dimensions reveal how methodological choices in dataset construction shape the reliability, scalability, and interpretability of experimental results.



 \textbf{Annotation Approaches (n=11)}: Several studies (P1, P6, P11, P12, P20, P52, P58, P70) used manual annotation performed by cybersecurity experts or trained annotators. Such annotation is used to tag entities and relationships, following established Named Entity Recognition (NER) schemes, such as BIO (P1 and P12). However, manual annotation is time and effort-intensive. To mitigate these challenges, several studies adopt semi-automated or hybrid annotation pipelines. For example, P1 combines dictionary-based matching using expert-curated entity lists with manual correction for out-of-vocabulary entities. P33 integrates static and dynamic malware analyses to generate comprehensive capability labels, while P54 leverages VirusTotal to automate malware labeling. Additionally, P62 employs dual labeling and overlap resolution strategies to handle sentences associated with multiple, interrelated techniques. Collectively, these approaches reflect ongoing efforts to balance annotation quality with efficiency.

\textbf{Annotation Tools and Frameworks (n=6)}: Several studies employed specialized annotation tools and frameworks to support structured labeling of CTI data. The BRAT annotation tool\footnote{https://brat.nlplab.org} is used in P6, P12, and P52 to facilitate manual annotation of entities and relations within CTI text. In contrast, P58 and P70 rely on custom-built, domain-specific annotation interfaces designed to capture complex relationships among adversarial entities, tactics, and indicators of compromise (IoCs). Beyond text-centric annotation, some studies operate on non-textual artifacts. P44 used Androguard\footnote{https://github.com/androguard/androguard} to extract control flow graphs (CFGs) from Android binaries, enabling program-level labeling of malicious behavior. Similarly, P25 employed CALDERA\footnote{https://github.com/mitre/caldera} to generate intrusion simulations that produce labeled Sysmon logs. These approaches extend annotation beyond natural language, reflecting the heterogeneous nature of CTI data.

\textbf{Quality Assurance through Inter-Annotator Agreement (n=4)}: Annotation reliability is explicitly addressed in four studies (P11, P57, P62, P70) through inter-annotator agreement (IAA) analysis. P11 reports agreement measurements between two independent annotators, while P57 adopts a multi-annotator protocol with explicit constraints on the number of labels assigned per document to improve consistency and reduce ambiguity. In P70, annotators receive prior knowledge on MITRE ATT\&CK concepts, and their annotations undergo multi-level review by additional cybersecurity researchers for further labeling reliability. These practices highlight the importance of quality assurance in mitigating subjectivity and ensuring dataset consistency.

 \textbf{Data Augmentation Techniques (n=6)}: To address data sparsity and class imbalance, five studies (P14, P35, P38, P43, P77)  employed data augmentation strategies.
 Lexical-level augmentation techniques are commonly used, with P43 and P77 adopting Easy Data Augmentation (EDA)~\cite{wei2019eda}, and P14 leveraging a Masked Language Modeling (MLM) based augmentation approach~\cite{devlin2019bert}. Beyond text-based augmentation, P35 applies AnyBURL\footnote{https://web.informatik.uni-mannheim.de/AnyBURL/}
 to infer additional triples and enrich existing cybersecurity knowledge graphs. More recently, P51 explores large language model (LLM)–based data generation, using GPT models\footnote{https://openai.com/open-models/}
 to synthesize CTI text samples. These approaches signal a shift toward generative augmentation techniques, although their effects on label correctness and annotation reliability remain unclear.
 
 \textbf{Ground Truth and Labeling Sources (n=6)}: Six studies (P1, P33, P45, P54, P55, and P79) derive labels from structured frameworks,most notably MITRE ATT\&CK. For example, P33 and P54 utilized sandbox analysis, VirusTotal metadata, and Virushare reports to automatically assign technique-level tags.On the other hand, P24 performed manual temporal and discourse-aware annotation on CTI corpora.
The authors manually annotated multiple datasets capturing (a) sentence-to-technique mapping, linking individual sentences to specific MITRE ATT\&CK techniques, (b) discourse relations between sentences (e.g., causal, conditional, elaborative), which contextualize how adversarial steps relate conceptually, and (c) temporal relations among ATT\&CK actions, encoding chronological dependencies. These efforts extend labeling beyond flat classification toward richer contextual modeling.

\textbf{Dataset Structuring and Representational Formats (n=7)}: Once annotated, datasets are often transformed into structured representations to support downstream modeling. Several studies (P12, P35, P58, P79) transform raw data into formal representations such as entity–relation triples, graphs, or paragraph-level multi-label datasets. For instance, P12 constructs entity–relation triples to support CTI relation extraction, while P35 enriches an existing APT knowledge graph by inferring additional triples using AnyBURL. P79 leverages Wazuh detection rules annotated with ATT\&CK technique identifiers to form structured rule–technique pairs, enabling downstream analysis of detection coverage. In other modalities,  P44 extracts control flow graphs (CFGs) from malware binaries to capture structural context associated with malicious behavior. Notably, a temporal trend emerges across the literature: earlier work predominantly relied on flat entity–relation representations (e.g., P12, P35), whereas more recent studies (e.g., P44, P45, P69) adopt hierarchical or graph-based structures that explicitly encode procedural, causal, and dependency relationships among attack components. This shift reflects growing recognition that richer structural representations are necessary to capture the complexity of adversarial behavior and support advanced reasoning over CTI data.

In summary, most datasets are manually annotated with MITRE ATT\&CK tactics or techniques, often at the sentence level. Some studies employ distant supervision or data augmentation to scale dataset construction, but annotation practices remain inconsistent across studies.

\section{TTPs Extraction Methodologies}\label{10}
In this section, we address \textit{RQ5: What methodological approaches have been employed for TTPs extraction?}
We analyze prior approaches for extracting TTP-related information from CTI data, focusing on text representation, modeling techniques for key extraction tasks, and the incorporation of domain-specific knowledge. The analysis outlines the progression from traditional feature-based methods to deep learning, including transformer and large language model (LLM) based approaches with cybersecurity-specific embeddings and fine-tuned architectures. The findings are presented in the following subsections.

\subsection{Representation of the CTI Text}
\label{textrepresentation}
In order to extract CTI from text using machine learning (ML), first, the text needs to be converted to a representation on which ML will be applied. Across the studies, we primarily observe three high-level categories of representations as given below.

\textbf{Traditional Representations (n=9):} Several works primarily relied on lexical and frequency-based representations, including TF–IDF vectors, Latent Semantic Indexing (LSI), and BM25. These representations were used for feature extraction (P20, P41, P63, P64, P73), semantic retrieval of attack techniques (P2, P3, P4), and ranking relevant techniques based on textual similarity (P64). While computationally efficient and interpretable, these methods are limited in their ability to capture context and semantic variation.

\textbf{Generic Embeddings (n= 21):} Since TFIDF-based representation cannot represent context and out-of-vocabulary situations, which are prevalent among cybersecurity texts, comparatively newer work relied on generic and stock embeddings, such as word2vec\cite{mikolov2013efficient} (P68), and transformers. BERT\cite{devlin2019bert}. BERT-based variants are the most widely used transformer-based embeddings, including RoBERTa, DistilBERT, Sentence-BERT, and XLM-Roberta. These embeddings are used mainly for named entity recognition for IoC extraction purposes (P11, P14, P26, P32, P52),  attack tactic and technique classification from text (P7, P19, P27, P47, P49, P55, P56, P59, P70, P79), and knowledge graph construction through entity and relation extraction (P13, P36, P52, P65), technique searching (P64). These representations provide contextual understanding and robustness to paraphrasing but remain limited by their general-domain pretraining.

\textbf{Custom Cybersecurity Specific Embedding (n=20):} Conventional BERT and BERT derivatives are trained on generic text, which do not understand cybersecurity domain-specific jargon. To overcome the limitation, several works obtained fine-tuned BERT models specifically trained on cybersecurity-specific texts. The examples of such models are CyBERT\cite{cybert2021}, SecBERT\cite{solanki2025secroberta}, SecureBERT\cite{aghaei2022securebert}, SecRoBERTa\cite{solanki2025secroberta}, Code-BERT\cite{feng2020codebert}, and SciBERT\cite{beltagy2019scibert}. These models are applied to technique classification (P10, P22, P42, P49, P55, P58, P76, P77), IoC extraction (P32), technique searching (P64), and knowledge graph construction (P11, P37, P60), and semantic role labeling (P80). In addition, several works fine-tuned their own BERT-based models. For example, P1, P5, and P75 fine-tuned their own BERT and RoBERTa models for IoC extraction and technique classification. P33 and P54 fine-tuned BERT models on API call graphs to classify techniques from system logs. Finally, P24 fine-tuned the RoBERTa model for extracting temporal relations among techniques. These efforts highlight the growing importance of domain adaptation in CTI modeling.

\subsection{Use of Natural Language Processing}

\textbf{Named Entity Recognition (n=15):}
Named Entity Recognition (NER) is widely used to identify cybersecurity-specific entities such as IoCs, malware names, tools, and actors.
CTI text contains cybersecurity-specific concepts, such as keywords, phrases, proper nouns, and code segments. The characteristics are suitable for applying named entity recognition (NER) on the text to extract these concepts. We observed NER for specific purposes: IoC extraction (P1, P11, P14, P16, P20, P26, P32, P73, P78), node classification in knowledge graph (P12, P29, P69), and relation extraction (P69) in the process of knowledge graph construction. Researchers primarily used TF-IDF and BERT in deep learning (i.e., LSTM, CNN, RNN)- based Conditional Random Fields (CRF) pipelines to process the text, while only one study (P51) explored LLM-based NER. Regular expression-based IoC extraction has been performed in P65, and attention-based architectures are explored in P37.

\textbf{Relation Extraction (n=16):}
To build a knowledge graph from CTI text, relation extraction is needed to infer relationships between nodes (entities) representing cybersecurity concepts. Most studies first apply NER to establish context and then infer relations using deep learning models, including graph neural networks (P13, P14, P24), CNNs (P37, P69), LSTMs (P12, P37, P69), traditional classifiers (P24), and LLMs (P6, P9, P17). BERT-based transformers have been used to represent the features of the node (P36, P52, P65). Several studies inferred relations among nodes via similarity measures: node similarity (P16), edge similarity (P61), and verb similarity (P61). Several natural language grammars have been utilized to infer potential relations, such as dependency parsing (P16, P23, P69), coreference resolution (P16, P23, P24, P61), subject-verb-object parsing (P23, P74), sentence positioning (P24), sentence distance (P24), shortest dependency paths (P65), static rules (P65), and the presence of temporal information related keywords (P24). 

\textbf{Text Classification (n=33)}
Text classification is a dominant approach for mapping CTI sentences to ATT\&CK techniques as mentioned in Section~\ref{papercategories}. For the classification task, researchers first converted text into an appropriate representation, as mentioned in the Section~\ref{textrepresentation}, and then fed the representation to the following types of learners: (a) traditional classifiers such as Naive Bayesian, Random Forest, Decision Tree, Gradient Boosted Trees, LinearSVC, Logistic Regression, K Nearest Neighbor and SVM (P20, P21, P22, P43, P53, P57, P59, and P79); (b) BERT, and BERT based variants (P5, P19, P24, P27, P59, and P80); (c) deep neural networks, such as CNN, RNN, LSTM, Siamese Network and Conditional Random Field (P22, P27, P38, P39, P42, P46, P50, P58, P59, P67, and P70); (d) LSI and BM25 ranking (P41, and P74) and (e) LLMs (P7, P28, P59, P71, P76, P77).

\subsection{Graph-Based and Structured CTI Modeling}
\textbf{Use of Heterogeneous Text Graphs (n=5):}
Several studies represent CTI as heterogeneous graphs to accommodate different types of nodes and edges. P13 builds a word--word/word--document/document--document graph to show the interrelation among CTI actors, malware, and attack techniques across different CTI reports. P14 and P47 construct an IOC-centric heterogeneous graph to relations among IoCs and techniques. Advanced graph algorithms have also been used to build the graphs, such as fuzzy matching (P47). and subgraph matching (P66). 

\textbf{Use of Knowledge Graph Embedding (n=3):} Knowledge graph embedding techniques are employed to infer missing relations and enrich CTI graphs.
Studies (P35, P47, and P54) used knowledge graph embedding techniques and negative sampling. P35 proposes a spatio-temporal knowledge graph embedding framework that strengthens embeddings by simultaneously enriching positive examples and generating high-quality negatives. P47 uses BERT-based contextual embeddings for entity initialization, applies fuzzy matching to align overlapping or conflicting nodes across heterogeneous CTI sources, and leverages a Graph Attention Network\cite{velivckovic2017graph} to propagate semantic dependencies within the graph. P54 proposes a multilayered knowledge graph embedding pipeline for mapping MITRE ATT\&CK techniques from malware source code, combining RoBERTa embeddings with GCN\cite{kipf2016semi}, TransE, and TransR\cite{asmara2023review} models, which can jointly capture both local structural dependencies and relational translation semantics.

\textbf{Temporal and Sequence Mining (n=4):}
We observe four studies that extracted temporal, dependency, and sequence-related information in the form of a knowledge graph instead of mining a STIX-based knowledge graph. For example, P24 builds a temporal knowledge graph using temporal relation classification (gradient boosting) using features from discourse relations (learned via next-sentence prediction on security text), technique co-occurrence, document position, lexical temporal cues, and mined recurring orderings. P16 and P17 extracted the dependency among techniques and assets via IOC replacement, NER, dependency parsing, coreference resolution, and manually curated technique templates. P61 extracts event sequences through text simplification, ellipsis handling, dependency events, and Snorkel\footnote{https://github.com/snorkel-team/snorkel} weak labels, then aligns nodes/edges with graph similarity.

\textbf{Clustering and Graph Analysis 
(n=5):}
Several studies leveraged clustering and graph-based techniques (P18, P25, P26, P30, and P67) to structure and interpret cyber threat intelligence (CTI) data by grouping similar techniques, behaviors, or actor profiles. Study P18 maps system log events to ATT\&CK techniques by converting textual procedure descriptions into rule-based command–technique mappings and then using K-means clustering with correlation-based similarity scores. P25 applied semantic clustering of attack techniques derived from honeypot logs, aiming to identify behavioral similarity across captured events. P26 performed clustering of APT actors using composite features such as MITRE ATT\&CK techniques, associated malware or software, and targeted industries. It utilized weighted similarity measures and hierarchical clustering to reveal latent organizational or campaign linkages. P30 applied social network analysis to an extracted knowledge graph to analyze structural relationships among CTI entities. P67 performed hierarchical clustering to categorize techniques based on learned latent representations. 

\subsection{Domain-Specific Adaptations}
\textbf{Use of Ontologies (n=4):}
Ontologies help sentence classification and knowledge graph extraction by providing structured semantic context that bridges raw text and domain knowledge. The most prevalent ontology that has been used in studies is STIX\footnote{https://stixproject.github.io/about/}, based on which the knowledge graphs are extracted. However, several studies performed technique classification and knowledge graph extraction based on their own-defined ontologies.  For example, P27 and P74 defined custom threat-action ontologies that specify semantic frames linking subjects, verbs, and objects, enabling the extraction of \textit{who did what to whom} patterns essential for TTP identification. In P55, the authors defined the COMAT ontology to guide semantic role labeling and verb–object pairing for the purpose of sentence technique classification. P36 integrated multiple established ontologies: CAPEC, CWE, and MITRE ATT\&CK, into a unified schema, for the purpose of cross-ontology reasoning and the inference of new relations among attack techniques, weaknesses, and mitigation strategies. P37 defined a domain ontology to formalize rule-based extraction of relationships among entities, improving both precision and interpretability in knowledge graph construction. 

\textbf{Text Normalization(n=6)}
Several studies performed different types of text normalization before the downstream CTI extraction task. P16 and P17 extract technique-dependency graphs via IOC replacement, entity and dependency parsing, coreference resolution, technique templates, and graph alignment for the extraction of the technique dependency graph. P61 adds text simplification, subject-ellipsis handling, and dependency events, and performs graph alignment using node, edge, and verb similarity to extract attack technique sequences. P70 catalogs 12 technique-specific element types (IPv4, domain, email, URL, hash, file path, regkey, CVE, crypto, protocols, filenames, data objects), replaces elements with placeholders, and performs technique classification based on the element co-occurrence. P8 proposes a rule-based system to resolve conflicts arising from the many-to-many relationships between IoCs and keywords. Furthermore, P40 uses semantic role labeling for IoC extraction purposes.

\textbf{System Log Classification (n=4):}
Although most of the studies focused on extracting CTI from technical reports, several studies mapped system logs and firewall rules into attack techniques. For the classification, static keyword-based rules (P18), BERT MLM (P33), CNN (P33), clustering (P18), graph neural networks (P78), and LLMs (P48) have been utilized to achieve the goal of the study.

\subsection{Use of Large Language Models (LLMs)}
Several studies (P6, P7, P17, P28, P71, and P77) have leveraged LLM-based pipelines for technique classification and knowledge graph construction. P6 experimented with a set of open-source LLMs, including LLaMA 2\footnote{https://huggingface.co/meta-llama} (7B–70B), Mistral 7B Instruct\footnote{https://huggingface.co/mistralai}, and Zephyr-7B-$\beta$\footnote{https://huggingface.co/HuggingFaceH4/zephyr-7b-beta}, using few-shot prompting and low rank adaptation (LoRA)-based fine-tuning to extract entities and relations for CTI knowledge graph construction. P7 conducted a comparative study of encoder and decoder-based LLMs (RoBERTa, SecureBERT, GPT-3.5-turbo) for tactic and technique classification, based on supervised fine-tuning, zero-shot inference, and retrieval-augmented generation (RAG) with FAISS\footnote{https://github.com/facebookresearch/faiss}-based nearest-neighbor retrieval to simulate real-world incomplete knowledge scenarios. Results indicate domain-adapted encoder models, such as SecureBERT, performed more consistently for supervised classification tasks, while GPT-3.5 demonstrated stronger generalization under zero-shot and RAG settings. P17, a follow-up to P16, leveraged LLMs to generate dependency-based attack graphs, incorporating IOC replacement, entity and dependency parsing, coreference resolution, and MITRE ATT\&CK template-based node alignment to improve contextual attack relationship modeling. P28 investigated LLM-driven technique identification with few-shot learning and retrieval-aware training (including reasoning-enhanced variants) using the following models: LLaMA 2 (2B–70B), Mistral 7B, GPT-3.5-turbo, and GPT-4. Findings suggest that larger instruction-tuned models such as GPT-4 and higher-parameter LLaMA variants performed better for complex reasoning tasks, particularly when identifying techniques embedded in long CTI narratives  

P71 focused on technique classification using LLaMA 2 models (7B, 13B, and 70B) and reported improved classification performance as model size increased, demonstrating that larger parameter models capture richer contextual representations of adversary behavior. Finally, P77 combined fine-tuned SecureBERT with LLM-based inference, following IOC replacement and data augmentation, demonstrating that hybrid pipelines integrating domain-specific encoders with LLM reasoning capabilities can enhance ATT\&CK-aligned classification accuracy.

Overall, these studies highlight a methodological transition from deep learning pipelines toward LLM-based and hybrid architectures. Domain-adapted encoder models (e.g., SecureBERT) tend to perform better for supervised classification on structured CTI datasets, while large generative models (e.g., GPT-4 or LLaMA-70B) demonstrate stronger performance in few-shot, reasoning-intensive, and retrieval-augmented settings. Collectively, the evidence suggests that combining domain-specific language models with LLM-driven reasoning and retrieval mechanisms offers the most effective approach for scalable CTI knowledge extraction and ATT\&CK technique classification.

\subsection{Weak Supervision, Data Augmentation, and Bias Correction}
Several studies leveraged weak supervision, data augmentation, and bias correction-based techniques due to the lack of a manually annotated dataset. For instance, P57 applied distant supervision using the Universal Sentence Encoder in combination with a support vector machine, a multi-layer perceptron, and a gradient boosting model. 61 adopts the Snorkel framework, defining labeling functions based on keywords, phrases, and IoC patterns to support entity extraction. To address dataset shift, P21 employed statistical reweighting techniques, including Kernel Mean Matching, the Kullback–Leibler Importance Estimation Procedure, Relative Density Ratio estimation, and confidence propagation~\cite{ayoade2018automated}. P58 used partial-ranking noise-contrastive estimation with asymmetric focusing to reduce gradient domination by easy negatives and to minimize mislabeled positives. P38 augments CVE descriptions using straightforward text augmentation strategies, while P39 explicitly models necessary and sufficient evidence to improve confidence in sentence-to-technique classification. Finally,  P10 develops a sentence-to-technique mapping dataset using active learning with a SciBERT classifier, using six sampling strategies such as top confidence, maximum entropy, margin sampling, Monte Carlo dropout, approximated gradient length, and core-set to efficiently prioritize informative samples. TTP extraction methods have evolved from rule-based and classical machine learning approaches to deep learning and transformer-based models, with recent studies increasingly exploring LLM-based extraction pipelines and knowledge-integrated frameworks for improved contextual understanding


\section{Evaluation Setup and Performance Metrics}\label{11}
In this section, we address \textit{RQ6: What metrics are used to evaluate performance?} We examine how prior work designs experimental evaluations, focusing on dataset usage, evaluation metrics, baseline selection, and qualitative validation practices. This analysis reveals common evaluation patterns and methodological limitations that affect the interpretability and generalizability of the reported results. The findings are discussed in the following subsections.

\subsection{Dataset Usage and Experimental Scope}
Most studies evaluate their proposed approaches using a single dataset (P1–P3, P5–P13, P15–P18, P20, P23–P28, P30–P38, P40–P46, P47–P57, P58–P63, P65–P68, P70–P80). While this practice simplifies experimental design and facilitates controlled comparison, it inherently limits the generalizability of results across different CTI sources, threat domains, and writing styles.
A small subset of studies conducts experiments across multiple datasets (P4, P14, P19, P21, P22, P29, P36), enabling limited cross-corpus validation. From the analysis, we observed that the predominance of single-dataset evaluations indicates that model transferability and robustness to distributional shifts remain insufficiently explored in the current TTP extraction literature.

\subsection{Evaluation Metrics}
We observe that the evaluation metrics primarily rely on precision, recall, and F1 scores. For technique and tactic classification, most work reports macro- and micro-averaged precision, recall, and F1-scores in multi-class or multi-label settings (P5, P11, P14, P15, P19, P24, P38, P39, P45, P47, P53, P55, P57, P67). 
Transformer-based encoder models generally achieve the highest macro- and micro-F1, followed by CNN/BiLSTM architectures, while TF–IDF and LSI-based classical models remain competitive baselines but degrade more sharply as the number of ATT\&CK technique labels increases. Several studies note that macro-F1 drops substantially under expanded label spaces due to class imbalance, even when micro-F1 remains relatively stable. To mitigate this, some works evaluate oversampling or cost-sensitive strategies, which improve minority-class recall and macro-F1 performance, particularly for traditional machine learning models.

A subset of studies adopts ranking-based metrics such as Label Ranking Average Precision, Hamming Loss, and Ranking Loss when modeling TTP extraction as a label-ranking problem (P19, P38). When framed as a retrieval or extreme multi-label task, information-retrieval metrics including Precision@K, Recall@K, HITS@K, Mean Reciprocal Rank (MRR), and Mean Average Precision (MAP) are commonly used (P6, P11, P24, P28, P35, P47, P50, P52, P58, P64). Across these works, semantic embedding–based retrieval approaches consistently outperform keyword-based systems, achieving higher top-K accuracy and ranking consistency.
Knowledge graph construction studies evaluate both entity and relation extraction using precision, recall, and F-measures (P6, P9, P12–P13, P16–P17, P29, P36–P37, P47, P51–P52, P61, P65, P69, P73). Entity extraction typically attains higher F1 scores than relation extraction, indicating that relational modeling remains the primary bottleneck. For graph completion and link prediction, embedding-based approaches are assessed using Hits@K and MRR (P6, P29, P35, P47, P52), and consistently outperform heuristic or rule-based completion strategies. For IoC extraction and named entity recognition, precision, recall, and F1 remain standard (P32, P40, P73, P78). However, only a small number of studies report entity-type–specific performance, limiting insight into per-indicator robustness. Clustering-based approaches employ unsupervised metrics such as Silhouette Score and Adjusted Rand Index (P26). 

Although TTP extraction tasks are inherently multi-class and multi-label with high label cardinality and severe imbalance, only a limited subset of studies provides rigorous multi-label evaluation, per-class analysis, or ranking-based assessment. In many cases, the evaluation design remains simplified relative to the problem's structural complexity, suggesting that methodological rigor in performance assessment lags behind the multidimensional nature of CTI extraction tasks.

\subsection{Baseline Comparisons and Benchmarking Practices}
Baseline comparison is a common practice across studies, particularly for technique and tactic classification. Frequently referenced systems include TTPDrill~\cite{husari2017ttpdrill}, TRAM~\cite{mitre_ctid_tram}, AttackG~\cite{li2022attackg}, and rcATT~\cite{legoy2019rcatt} (P8, P11, P15, P16–P17, P19, P51–P52, P55, P61, P64, P69, P75, P77). Beyond prior CTI-specific systems, studies benchmark against a broad range of methodological baselines, which can be grouped into the following categories:


\begin{enumerate}
    \item Deep Learning Models: BERT, RoBERTa, SecureBERT, SciBERT, DistilBERT, XLM-RoBERTa (P5, P19, P22, P27, P32, P37, P41–P42, P49, P51–P52, P55, P57, P67, P70)

    \item Large Language Models: GPT-3.5/4 and open families such as LLaMA, Qwen\footnote{https://qwen.ai/home}, Baichuan\footnote{https://github.com/baichuan-inc/baichuan-7B/blob/main/README\_EN.md}, Mistral (P7–P8, P28, P48, P51, P64, P71)
    
    \item Multilabel Classification Baselines: Binary Relevance, Label Powerset, Classifier Chains (P41, P43, P53).

    \item Traditional ML Models: SVM, Random Forest, Decision Tree, Naive Bayes, MLP (P3, P11, P15, P21–P22, P41–P46, P54, P57, P59, P66, P67, P70, P76).
    
    \item Data Augmentation: Easy Data Augmentation, WordNet, active learning approaches (P10, P14).

    \item Graph Neural Network: GCN, TextGCN, BertGCN, GAT, GIN (P13–P14, P44, P54, P72)
    
    \item Traditional Approaches for NER and RE: CRF, BiLSTM-CRF, CNN-BiLSTM-CRF, BERT-LSTM-CRF, BI-GRU, BI-LSTM with attention (P1, P12, P13, P23, P27, P37, P51, P52).
    
    \item Graph Neural Networks and Embeddings: TransE, TransH, DistMult, ComplEx, Relgraph, AGGCN (P6, P29, P35, P47, P52).

\end{enumerate}

\subsection{Qualitative Evaluation and Case-Based Analysis}
Several studies incorporate qualitative evaluation through case studies and explainability analyses. These studies include feature attribution and explanation techniques such as LIME and SHAP (P39, P80), entity or tactic-level error analysis (P9, P24, P77), and detailed discussion of real-world prediction failures and their underlying causes (P8, P56, P61). Such case-wise or class-wise reporting is most commonly reported in NER and multi-label classification settings (P24, P32, P52). However, systematic and standardized interpretability evaluation remains limited, underscoring the need for more principled assessment of model transparency and failure modes in CTI extraction.


\section{Replicability}\label{12}
The data availability across the selected papers is not always consistent or comprehensive. Across the 80 reviewed papers, we observed varying levels of replicability and data availability, with significant implications for the reproducibility and generalizability of TTP extraction research.

While 47.5\% (N=38, P4, P6, P7, P10, P11, P16, P22, P23, P24, P25, P28, P31, P36, P37, P38, P41, P43, P46, P49, P52, P53, P55, P56, P57, P58, P60, P61, P62, P64, P65, P66, P68, P70, P71, P73, P75, P77, P79) of the papers provide at least one public dataset or repository, either through datasets, code repositories, only 12.5\% (n=10, P11, P22, P36, P38, P43, P55, P58, P64, P66, P75) make both the code and data publicly available, which significantly limits reproducibility. These resources were typically hosted on platforms such as GitHub, GitLab, Figshare, Zenodo, or Kaggle, or were described as publicly available in the papers.

A substantial portion, 20.0\% (n=16,  P6, P7, P10, P16, P23, P25, P28, P37, P41, P52, P61, P62, P65, P70, P71, P77), provide code repositories through platforms such as GitHub or GitLab, but do not explicitly confirm that accompanying datasets are accessible, leaving questions about complete reproducibility. Another 15.0\% (n=12, P4, P24, P31, P46, P49, P53, P56, P57, P60, P68, P73, P79) mention dataset availability or describe data as publicly accessible but do not provide implementation code, which may hinder validation of their proposed approaches.

Several papers rely on proprietary, confidential, or manually curated datasets that are either unavailable or only accessible upon request, making independent verification challenging. Specifically, 50.0\% (n= 40, P1, P2, P3, P8, P9, P12, P13, P14, P15, P17, P18, P19, P20, P21, P26, P27, P29, P30, P32, P33, P34, P35, P39, P40, P42, P44, P45, P47, P48, P50, P51, P54, P63, P67, P69, P72, P74, P76, P78, P80) do not clearly report dataset or code availability, making it difficult to determine whether reproduction would be feasible without directly contacting authors. Furthermore, only 2.5\% (n=2, P5, P59) explicitly state that resources are unavailable, with one case (P59) mentioning a GitHub repository that contains no actual data despite claims of availability. Moreover, we observed inconsistencies in how papers describe their datasets and experimental configurations. Some report high-level statistics such as the number of threat reports analyzed or sentences extracted, while others, approximately 35-40\% of the studies with available resources, did not provide sufficient detail on annotation procedures, data cleaning steps, train-test splits, or hyperparameter settings. These discrepancies limit the ability to evaluate the reliability and applicability of findings across the broader CTI extraction community and make direct comparison between approaches challenging. Additionally, concerns arise regarding the long-term viability of shared resources. Several papers reference links that may become inactive over time or point to repositories with incomplete documentation. Without proper versioning, DOIs, or archival practices, the sustainability of resources remains uncertain. This is particularly problematic given the evolving nature of CTI sources. Updated threat databases like MITRE ATT\&CK and emerging attack techniques mean that many datasets lack versioning or timestamps, affecting long-term reproducibility and relevance.

These findings highlight a significant reproducibility gap in TTP extraction research. The relatively low proportion of papers providing both code and data (12.5\%), combined with the high percentage offering unclear or no availability information (50.0\%), underscores the need for stronger reproducibility standards, more comprehensive resource sharing practices, and clearer reporting guidelines in the field.

\section{Reported Limitations}\label{13}
Many of the reviewed papers explicitly mention limitations in their study designs, primarily related to data coverage, annotation, model scope, and evaluation. A commonly stated limitation is the restricted scope or size of the dataset used in several papers, acknowledging that their datasets focus on specific APT groups, regions, or time frames, which may limit the generalizability of their findings. Some papers note that their models are only evaluated on English-language reports and may not perform well in multilingual or non-English threat contexts. Additionally, multiple papers highlight the lack of ground truth in threat intelligence data as a significant challenge, especially for evaluating TTP extraction systems. In response, some studies rely on heuristics or distant supervision techniques, but also recognize that these methods introduce label noise and potential inaccuracies.

Several papers report that their annotation processes are based on expert knowledge but admit they do not include inter-rater agreement scores, making it difficult to assess labeling reliability. Other studies mention challenges in clearly defining TTP boundaries or classifying vague or implicit threat behaviors, which complicate both annotation and extraction. Papers that employ rule-based or pattern-matching techniques often state that their approaches may not generalize well to unseen data or novel attack techniques. Some deep learning-based papers note limitations such as small sample sizes for training, leading to potential overfitting, or the absence of fine-grained entity-level supervision. A few papers also reflect on the limitations of their evaluation settings, for example, by acknowledging that their methods were only tested in simulated or static environments and not validated against real-time or evolving threat data. Overall, these reported limitations highlight the complexity and evolving nature of CTI extraction and point to the need for more comprehensive, scalable, and generalizable approaches in future research.

\section{Threats To Validity}\label{14}

As with any systematic review, this study is subject to several threats to validity. First, the search strategy relied on five major scholarly databases; as a result, relevant studies from preprint servers, practitioner venues, or non-indexed repositories may have been missed, potentially introducing selection bias. To mitigate this risk, we applied backward and forward snowballing on the initially selected papers and conducted multi-round screening by two independent reviewers using predefined inclusion and exclusion criteria. Second, qualitative synthesis of TTP extraction methodologies involves researcher interpretation, which may introduce coding bias when categorizing studies by task, methodology, or evaluation practices. To reduce this risk, two authors independently coded each paper following a structured extraction protocol, resolved disagreements, and assessed inter-rater reliability using Cohen’s kappa. Finally, the reviewed studies exhibit variation in datasets, evaluation metrics, and experimental setups, particularly for multi-label tasks such as TTP extraction and IoC recognition. This heterogeneity limits direct quantitative comparison across studies. Therefore, we adopted a qualitative, task-aware synthesis that emphasizes methodological trends and evaluation practices rather than aggregating performance results numerically.

\section{Future Research Direction}\label{15}
Despite significant advances in automated extraction of TTPs from CTI, we found several persistent gaps that limit the operational impact of existing approaches. In this section, we recommend and outline key directions for future research that can strengthen the empirical foundations of TTP extraction and better align methodological advances with practical cybersecurity needs.

\textbf{Grounded datasets from operational CTI:} A major limitation in existing work is the reliance on curated, filtered, or synthetic datasets that fail to capture the noise, ambiguity, and stylistic diversity present in operational CTI reports. As a result, many proposed methods are evaluated under simplified conditions that may not reflect real-world threat intelligence environments. Future research should therefore prioritize the creation and release of high-quality, publicly available datasets derived from practitioner-generated CTI reports. Such benchmark datasets would enable more realistic evaluation, facilitate reproducibility, and significantly improve the external validity of TTP extraction approaches.

\textbf{Multi-label evaluation paradigms:}
Although CTI narratives frequently describe multiple tactics and techniques within a single report, many studies continue to evaluate TTP extraction using single-label or simplified classification settings. Future work should adopt multi-label learning and evaluation paradigms that explicitly model overlapping, co-occurring, and hierarchically related TTPs. Such paradigms would better reflect adversarial behavior and provide more faithful assessments of model performance.

\textbf{Granular performance reporting:}
The literature largely emphasizes aggregate metrics such as overall precision, recall, or F1-score, with limited reporting of per-class, per-technique, or per-document performance. Future studies should report fine-grained metrics that expose variability across TTP categories, data sources, and contextual settings. Granular reporting would enable deeper diagnostic analysis, support fairer comparison across methods, and reveal systematic strengths and weaknesses that are obscured by aggregate scores.

\textbf{Beyond STIX-centric knowledge representations:}
Most knowledge graph–based approaches rely on STIX-oriented representations, which primarily encode structured indicators and static relationships. While valuable, these representations often undercapture temporal ordering, procedural dependencies, causal relationships, and evolving attacker intent. Future research should explore richer, narrative-aware representations that integrate temporal, causal, and contextual dimensions extracted directly from CTI text to support more expressive reasoning about adversary behavior.

\textbf{CTI-driven adversary emulation:}
An emerging research direction involves leveraging CTI to support adversary emulation and attack simulation. In particular, large language model–based systems offer the potential to generate realistic attack sequences or playbooks grounded in observed threat behavior. However, rigorous evaluation is needed to assess the fidelity, security risks, and operational utility of such approaches, especially in safety-critical or defensive contexts.

\textbf{Context-Aware TTP Extraction:}
Many existing approaches perform TTP extraction at the sentence or token level, often treating each sentence independently. However, CTI reports frequently describe attack behaviors through multi-sentence narratives, where important contextual cues are distributed across paragraphs or sections of the report. As a result, sentence-level models may fail to capture implicit relationships, the temporal ordering of attack steps, or dependencies between tactics and techniques. Future research should therefore explore context-aware extraction models that incorporate broader document-level information.


\textbf{Model generalization and robustness:}
A few studies examine the robustness and generalization of TTP extraction models across datasets, threat domains, or time periods. Future work should incorporate cross-dataset validation, temporal evaluation, and robustness analyses to assess susceptibility to underfitting, overfitting, concept drift, and domain shift. Such analyses are essential for understanding deployment readiness and long-term reliability in evolving threat landscapes.

Addressing these directions would advance the current state of TTP extraction research and enhance its applicability to real-world threat analysis, detection, and defensive operations.

\section{Conclusion}\label{16}

 TTPs can be automatically extracted from unstructured CTI texts produced by security vendors, researchers, and threat analysts to support understanding of adversary behavior. In this systematic literature review, we analyzed 80 peer-reviewed studies on TTP extraction from textual CTI sources. We organized the literature along multiple dimensions, including extraction purposes, data sources, data collection and preprocessing strategies, dataset annotation and construction practices, and applied methodologies, to provide a structured view of the research landscape. Our analysis identifies five major categories of TTP extraction objectives addressing different threat intelligence goals and nine categories of data sources, with CTI reports and security advisories being the most frequently used. We further observe ten categories of data collection and preprocessing techniques, and six categories each for dataset annotation strategies and methodological approaches. Across these dimensions, the literature predominantly focuses on technique-level classification, typically using supervised learning models trained on sentence-level datasets derived from CTI reports or ATT\&CK descriptions. While earlier studies relied on rule-based methods and traditional machine learning approaches, more recent work increasingly adopts transformer-based architectures and LLM–driven pipelines, reflecting a broader shift toward contextual and knowledge-aware extraction techniques. Nonetheless, several limitations remain in the current research landscape. Many studies rely on narrowly scoped datasets derived from limited CTI sources; evaluation practices frequently report aggregate performance metrics without cross-dataset validation; single-label or simplified classification settings are used for evaluation; and the availability of dataset transparency and artifact availability remains inconsistent, hindering reproducibility and systematic benchmarking. To address these limitations, future research should focus on developing more transparent and reusable datasets, richer annotation schemes supporting multi-label and contextual TTP relationships, and evaluation frameworks that better reflect real-world CTI workflows. Advancing these directions will be valuable for building more scalable, reproducible, and operationally useful TTP extraction systems. Overall, this work consolidates existing knowledge, identifies key gaps, and outlines a research roadmap to advance automated TTP extraction in support of more robust and operationally useful CTI-driven security analysis.


\newpage


\bibliographystyle{ACM-Reference-Format}
\bibliography{main}

\appendix

\end{document}